\pdfoutput=1 

\documentclass[12pt]{iopart}
\usepackage{graphicx}
\usepackage{amssymb}
\usepackage{amsbsy}

\usepackage{iopams}
\usepackage{hyperref}

\newcommand{\beq}{\begin{equation}}
\newcommand{\eeq}{\end{equation}}
\newcommand{\be}[1]{\begin{equation}\label{#1}}
\newcommand{\ee}{\end{equation}}
\newcommand{\continue}{\nonumber \\ }
\newcommand{\bea}{\begin{eqnarray}}

\newcommand{\eea}{\end{eqnarray}}

\newcommand{\bm}[1]{\mbox{\boldmath{$ #1$}}}

\renewcommand\Im{{\rm Im\,}}

\newcommand{\refeq}  [1] {(\ref{#1})}

\newcommand{\scatmat}{\mathbb{S}}

\newcommand{\atan}{\mathrm{atan}}
\newcommand{\acos}{\mathrm{acos}}
\newcommand{\xx}{{\Delta x}}
\newcommand{\axx}{{|\xx|}}

\begin{document}

\title{Families of spherical caps: spectra and ray limit}
\author{Niels S{\o}ndergaard\dag \, and Thomas Guhr\dag\P}
\address{\dag\ Division\, of Mathematical\, Physics, Lund University, LTH, Sweden}
\address{\P\ Fachbereich Physik,  \, Universit\"{a}t \,   Duisburg-Essen, Germany}

\begin{abstract}
We consider a family of surfaces of revolution ranging between a disc
and a hemisphere, that is spherical caps. For this family, we study
the spectral density in the ray limit and arrive at a trace formula
with geodesic polygons describing the spectral fluctuations.  When the
caps approach the hemisphere the spectrum becomes equally spaced and
highly degenerate whereas the derived trace formula breaks down. We
discuss its divergence and also derive a different trace formula for
this hemispherical case. We next turn to perturbative corrections in
the wave number where the work in the literature is done for either
flat domains or curved without boundaries.  In the present case, we
calculate the leading correction explicitly and incorporate it into
the semiclassical expression for the fluctuating part of the spectral
density. To the best of our knowledge, this is the first calculation
of such perturbative corrections in the case of curvature and
boundary.

\end{abstract}

\section{Introduction}

Waves can be approximated by rays by means of asymptotic expansions.
For example in the design of optical lenses one always resorts to the
equations describing the rays of light rather than the Maxwell
equations for the electromagnetic waves. Similarly for the
interpretation of earthquakes, one replaces elastic waves with curved
rays. 

The parameter of such an expansion for a classical wave phenomenon is
the wave length divided by a typical geometrical scale of the
systems. In optics, this leads from electromagnetic waves to rays of
light.  In quantum mechanics, however, the asymptotic parameter is
$\hbar$ compared to a typical action of the system, that is, it
involves a natural constant.  Nevertheless on the formal mathematical
level, the asymptotic expansions have the same character and many
semiclassical results have their analogue for classical wave equations
in terms of similar ray interpretations.  Likewise, there are trends
in the opposite direction, from classical waves to quantum mechanics,
such as imaging by electron waves in e.g. photodetachment microscopy
\cite{BD06}.  For this article we stress that the semiclassical
discussion of a free quantum mechanical particle on a curved surface
is similar to that of the ray theory for a classical membrane. Thus,
in curved spaces, the effect of geometry on quantum mechanics is that
wave packets follow {\it geodesic motion} in the semiclassical limit
in the absence of potentials \cite{Fried75}. Geodesics are curves
which are stationary with respect to the length under variations.

In this contribution we consider a one-parameter family of spherical
caps. We show that detailed information about the spectra is obtained
from periodic geodesic orbits.  In particular, we explain the drastic
change in the spectrum involving a clustering effect when approaching
the half-sphere by varying the opening angle of the cap.  
Besides the method of stationary phase, we shall use the asymptotic
technique attributed to Jeffreys, Wentzel, Kramers and Brillouin and denote
it {\it JWKB}. We shall discuss how to take into
account the effect of the boundary with respect to the first
JWKB-correction using the scattering approach of
\cite{smilanskyCourse}.

Our interest in spherical caps is motivated by the ongoing
experimental efforts for elastic shell caps  by  Ellegaard and his group.

\begin{figure}[ht]
\label{fig:NotationCap}
\begin{center}
\includegraphics[viewport=88 4 520 328.7,scale=0.7]{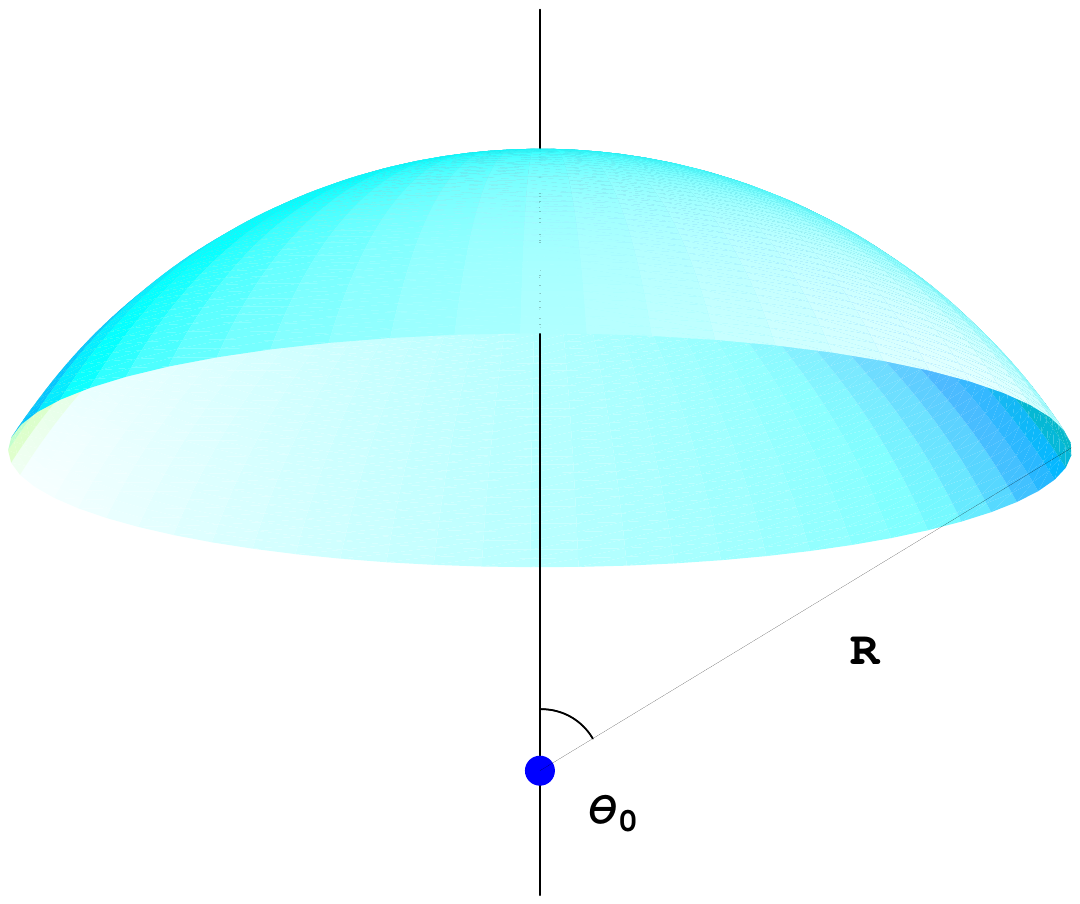}
\caption{The cap geometry: opening angle $\theta_0$ and curvature radius $R$}
\end{center}
\end{figure}

The article is organized as follows: to aid a classical
interpretation we  recall various facts
about geodesic polygons in \sref{sect:Classical}. Then the spectral problem is introduced in
\sref{sec:waveIntro} with a discussion of the asymptotics of the
associated Legendre polynomials.  Using a scattering approach to
quantization, this allows a derivation of a trace formula for the
density of states, see \sref{sec:ScatPhase}. Then follows a discussion
of the pure hemisphere in \sref{sec:Hemi} which can be solved exactly
along with a comparison in \sref{sec:CompareTrF} with the previous
general case.  We include JWKB-corrections in the trace formula in
\sref{sec:JWKBCorr}. We end with a short discussion in
\sref{sec:conclusion}. Finally, calculational details are given
in appendices.

\section{Classical quantities: geodesic polygons}
\label{sect:Classical}
We  fix the notation in \sref{sec:FixNot} and discuss the conservation of angular momentum during the geodesic flow in \sref{sec:Lcons}. For reference later on, we compile  a list of geometric quantities for geodesic polygons in \sref{sec:OrbQuant}, give a condition for the  inclination of a segment in a geodesic polygon in \sref{sec:OrbCond} and discuss the  concept of anholonomy present in systems with continuous symmetries such as $U(1)$  with a calculation for the case of the cap in \sref{sec:Anhol}.

\subsection{Notation}
\label{sec:FixNot}
We shall consider a family of surfaces of revolution. 
Individual members of this family are spherical caps and  are parameterized by a radius of curvature $R$ and an opening angle $\theta_0$ calculated from the axis of symmetry taken to be the $z$-axis, see \fref{fig:NotationCap}. We consider caps ranging from almost a disc to a hemisphere and we restrict ourselves to caps with opening angles less than or equal $\pi/2$. The singular cases corresponding  to $\theta_0 =0,\pi$   are not considered. 

\subsection{Conservation of angular momentum}
\label{sec:Lcons}
If a surface possesses
symmetries these  influence the dynamics of the geodesic flow. 
For example,  for surfaces of revolution Clairaut
\cite{clairaut1733} found from the geodesic equations: \beq
\label{eq:Clairaut} r \, \cos \Theta = \mathrm{Cst.} \, , \eeq where $r$
is the radial distance from the rotation axis and $\Theta$ is the
angle of the tangent vector of the geodesic with a given latitude,
see \cite{spivak}.  When \refeq{eq:Clairaut} is multiplied with the
momentum $p$, the corresponding physical interpretation is that the
angular momentum along the rotation axis is conserved. Here, one uses
the fact that the momentum is conserved in the absence of potentials
\cite{Fried75}. 

In the presence of a boundary, the geodesic flow is broken but it shall 
be continued by allowing geodesics to follow the law of
reflection. Therefore, for the cap geometry the periodic orbits will
be geodesic polygons composed of {\it great circles}. In the present
case, the reflections are operations which preserve angular momentum.

Likewise, from a spectral point of view, the angular momentum is a
good quantum number.

\subsection{Geometrical orbit quantities}
\label{sec:OrbQuant}
\begin{figure}
\begin{center}
\includegraphics[viewport=88 4 520 381, scale=0.5]{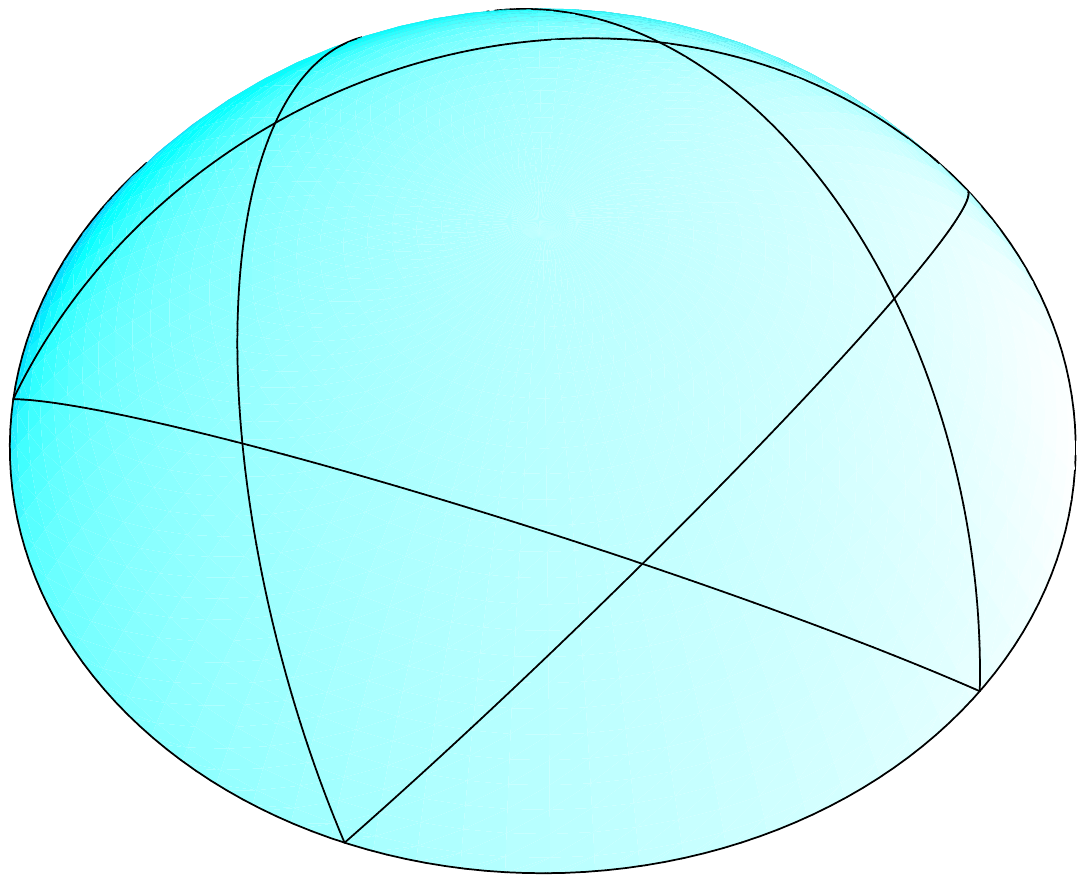}
\caption{\label{pentagram} The pentagram orbit for opening angle $\theta_0=\pi/3$.}
\end{center}
\end{figure}

We label each geodesic polygon  with  $n$ edges and winding number $M$ \footnote{i.e. how many times the orbit winds around the north pole}
 with a pair of numbers $(n,M)$. The pentagram orbit for example has $(n,M)=(5,2)$, see \fref{pentagram}. Consider a single edge in such a polygon. Let its vertices be given in direction by the two unit vectors 
$\mathbf{e}_1 $ and $\mathbf{e}_2$ with 
\beq
\label{eq:E1vec}
\mathbf{e}_1 = \sin \theta_0 \, \mathbf{\hat{x}} +\cos \theta_0 \,
\mathbf{\hat{z}} \eeq and 
\beq 
\label{eq:E2vec}
\mathbf{e}_2 = \sin \theta_0 \cos
\delta \phi\, \mathbf{\hat{x}} + \sin \theta_0 \sin \delta \phi \,
\mathbf{\hat{y}} + \cos \theta_0 \, \mathbf{\hat{z}} \, ,
\eeq 
where the
azimuthal angle traversed is  denoted by 
\beq \label{def:azimIncr} \delta
\phi = \frac{2 \pi M}{n} \,.  \eeq 
So the angle from the origin between the two points defining the segment
\beq \label{eq:SegmentAngle}
\chi \equiv \angle(\mathbf{e}_1,\mathbf{e}_2)  
\eeq 
can be found by e.g.
\beq \label{eq:E1DotE2}
\cos \chi 
=\mathbf{e}_1 \cdot \mathbf{e}_2 = \cos^2 \theta_0 + \sin^2 \theta_0 \cos \frac{2 \pi M}{n} \,.
\eeq 
Hence the length of this
polygonal edge is
\beq \label{eq:LengthSegment}
\Delta l = R \, \chi \,  \eeq 
corresponding to one $n$'th of the total length of the orbit.
\subsection{Orbit inclinations from conservation of angular momentum}
\label{sec:OrbCond}
With respect to the origin a
particle moving on the segment with momentum $p$ has an angular
momentum $\mathbf{L}$.  Classically $|\mathbf{L}|$ and $L_z$ are
conserved. The latter is also conserved under reflections from the
boundary.   The direction becomes \beq
\mathbf{\hat{L}} = \frac{\mathbf{e}_1 \times
\mathbf{e}_2}{|\mathbf{e}_1 \times \mathbf{e}_2|} \eeq and the
magnitude is \beq |\mathbf{L}| = R p \,.  \eeq 
Using \eref{eq:E1vec} and \eref{eq:E2vec}, we find after some calculation
the projection of the angular momentum  on the
$z$-axis: 
\beq 
\label{eq:LzClass}
L_z  = p  R \, \cos \psi 
\eeq with 
\beq
\label{eq:cosClass}
\cos \psi = \frac{\sin \theta_0 \cos \frac{\pi M}{n}}{\sqrt{\cos^2 \theta_0 + \sin^2 \theta_0 \, \cos^2 \frac{\pi M}{n}}} \,.
\eeq
In the following, we shall interpret \refeq{eq:cosClass} as a condition fulfilled by an orbit's inclination (measured relative the polar axis) when given winding number $M$, number of bounces $n$ and the opening angle of the cap $\theta_0$.

\subsection{Anholonomy}
\label{sec:Anhol}
We consider the geodesic flow as a dynamical system. We shall
 record the position of the trajectory at the boundary by a
 corresponding azimuthal angle $\phi$ and think of it as a phase. For
 an initial incidence angle this phase will change by constant
 discrete increments $\delta \phi$ and only if this phase is given by
 \refeq{def:azimIncr} the orbit is closed. We now enquire how this
 phase changes as the direction of the orbit changes. This direction
 we choose to be controlled by $L_z$, the classical variable conjugate
 to the azimuthal variable $\phi$.  Thus, if a general azimuthal
 increment $\delta \phi$ changes, $L_z$ changes via \refeq{eq:E2vec},
 \refeq{eq:E1DotE2} and \refeq{eq:LzClass}.  After some calculation
 this yields \beq
\label{eq:PhaseLag} \frac{\partial \delta \phi}{\partial L_z} = -
\frac{2}{p R} \, \frac{(\cos^2 \theta_0 + \sin^2 \theta_0 \, \cos^2
\left({\delta \phi \over 2}\right) )^{3/2}}{\sin \theta_0 \cos^2
\theta_0 \sin \left( {\delta \phi \over 2} \right) } \, \eeq the {\it
phase lag} \cite{CreaghLittle,Creagh}, also called the anholonomy.

\section{Semiclassics using asymptotics of Legendre polynomials}
\label{sec:waveIntro}
We have finished the classical considerations and will next study spherical caps using wave mechanics.  First, we shall discuss the wave problem of a spherical surface  and the corresponding symmetry reduced problem given by Legendre's equation in \sref{sec:PlmSetup}. We find the ray-limit of  this reduced wave equation  using the JWKB-method in \sref{sec:JWKBLeg}. This will lead to asymptotics for the associated Legendre polynomials. 
\subsection{Wave problem of scalar cap}
\label{sec:PlmSetup}
 The simplest model of cap vibrations corresponds to that of a quantized particle confined in the cap region:
\beq \label{curvHelmholtz}
(\Delta + k^2) \Psi = 0
\eeq
with $\Delta$ the curved Laplacian, k the wave number and specified boundary conditions such as Dirichlet $\Psi = 0$ or Neumann $\partial_n \Psi =0$ at $\theta=\theta_0$. \Eref{curvHelmholtz} is also considered to describe curved drums. 

\begin{figure}[ht]
\begin{center}
\includegraphics[viewport= 88 4 520 271,scale=0.7]{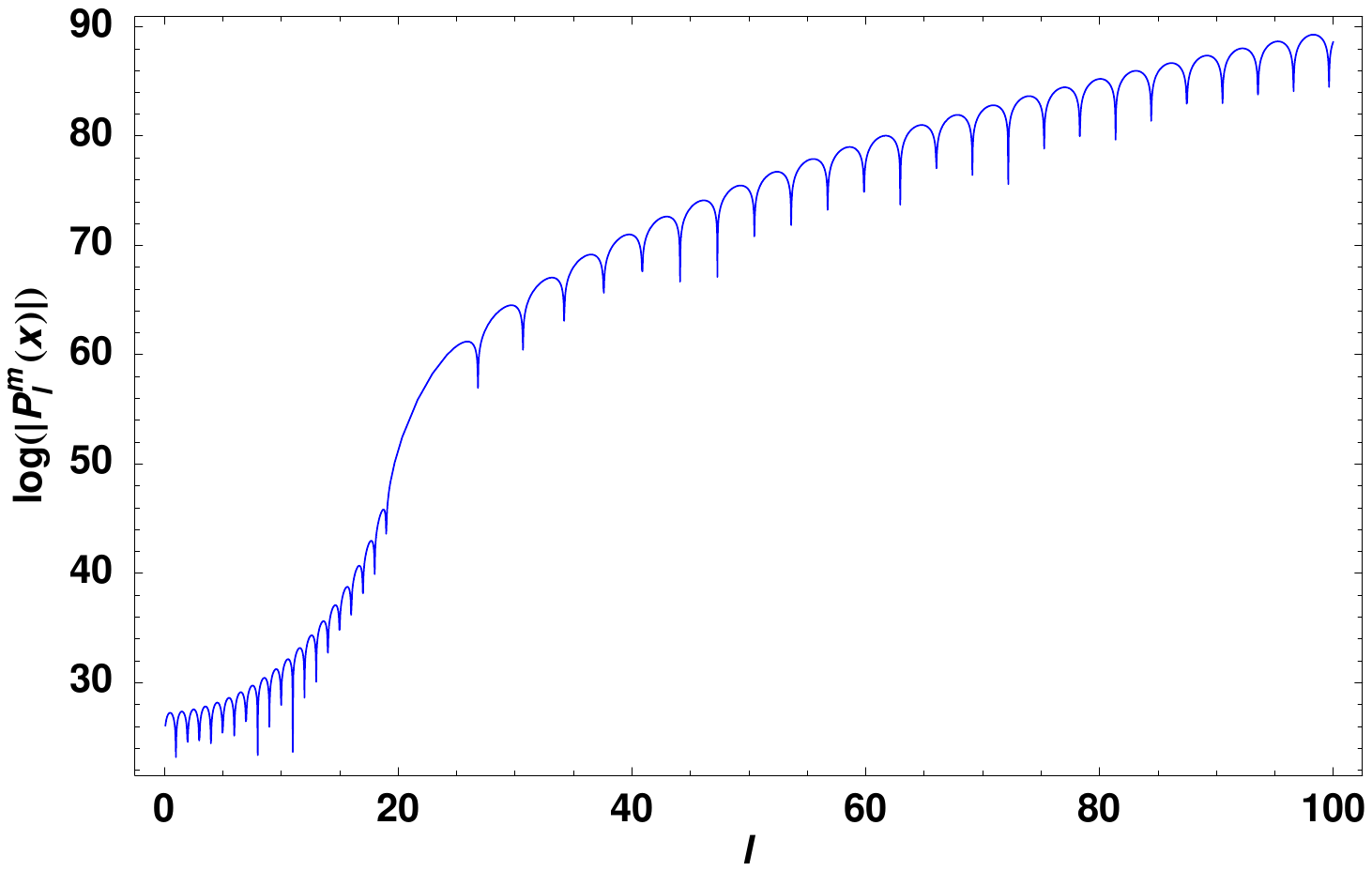}
\caption{\label{figSubSpec} The values of $\log |P_l^m(x)|$ for $m=20$ and
$x=\cos( \pi/3)$. For $l>m$ the dips correspond to eigenmodes being roots of
$P_l^m(x)=0$, see \sref{sec:PlmSetup}. }
\end{center}
\end{figure}

When solving \refeq{curvHelmholtz} by separation of variables using
\beq \Psi = u(\theta) e^{i m \phi} \eeq Legendre's equation \beq
\label{legendreODE} (1-x^2) \, \frac{d^2 u}{dx^2} - 2 x \,
\frac{du}{dx} +\left(l(l+1) -\frac{m^2}{1-x^2} \right)u = 0 \, \eeq
arises with the variable $x= \cos \theta$ and the parameter \beq
l(l+1) = (k R)^2 \equiv \kappa^2 \, \eeq related to the spectrum.  The
regular solutions at $\theta=0$ are given by the associated Legendre
polynomials $P_l^m(x)$ whereas the irregular ones by
$Q_l^m(x)$. Consequently the Dirichlet solutions are the solutions of
\beq
\label{eq:QuantCond}
P_l^m(x_0)=0 \eeq with respect to $l$ and where $x_0\equiv \cos
\theta_0$, see \fref{figSubSpec} showing $\log(|P_l^m(x)|)$ in the
case $m=20$ and $\theta_0=\pi/3$. When $l<m$ the dips show an integer
spacing. There as $m$ is an integer $P_l^m(x) = 0$ for all $x$ and
integer $l<m$, since in general at integer $l,m $ the $P_l^m$ is
proportional to an $m$'th derivative of an $l$'th order
polynomial. These ``roots'' are indeed present in \fref{figSubSpec} up
to $l=m-1=19$ but do not correspond to genuine eigenmodes as the wave
function is globally zero.  Around $l\approx m$ there is a gap after
which physical states appear spaced apart typically with a non-integer
value for high $l$. The reason for this fixed spacing is due to the
following asymptotics when $l \gg m$  \cite{abramov}:
\beq \fl P_l^m(\cos \theta) = \frac{\Gamma(l+m+1)}{\Gamma\left(l+{3 \over 2}\right)} \,
(\pi \sin \theta)^{-1/2} \, \cos\left(\left(l+{1 \over 2} \right) \theta -{\pi \over 4} +m {\pi \over 2}
\right) +O(l^{-1}) \,.  \eeq To further understand the roots of
condition \refeq{eq:QuantCond} in the interface region where $l$ and
$m$ are comparable we turn to a more detailed asymptotics
of the Legendre polynomials.

\subsection{JWKB of Legendre polynomials}
\label{sec:JWKBLeg}
Standard tables of functions often discuss the asymptotics of
$P_l^m(x)$ in the extreme cases where either $l \gg m$ or $m \gg l$.
However, as mentioned we need the more interesting intermediate
physical situation with both $l$ and $m$ large and their ratio well
defined following \cite{edmonds} (more explanatory details can be
found in \cite{BPC96, DT98, BT57} ).

Thus, consider large angular momenta $l$ and from that define a small parameter
\beq \label{eq:Epsilon}
\epsilon = (l (l+1))^{-1/2} \equiv \frac{1}{k R} \,
\eeq
and demand that the $z$-component of the angular momentum $m$ to be of the same asymptotic order as $l$. Hence, define the projected angular momentum on the $z$-axis respective $x$-$y$-plane
\beq 
m= L_z  \qquad \mathrm{and} \qquad  l_\perp =\sqrt{l (l+1) - m^2 } \,.
\eeq
The classical picture is the precession of the angular momentum around the $z$-axis at the  angle $\psi$, where we defined the corresponding cosine  in \eref{eq:cosClass}. Thus  for semiclassics we shall associate a classical angle $\psi$ to the quantum mechanical angular momentum by
\beq \label{eq:AorPsi}
\cos \psi = \epsilon \,  m \qquad \mbox{and} \qquad \sin \psi = \epsilon \, l_\perp  \equiv a \,.
\eeq
To simplify the following analysis,  switch from the original wave function $u$ to
\beq
\label{eq:NewWaveFct}
w(x) \equiv \sqrt{1-x^2} \, u(x) \, .
\eeq
Then Legendre's equation \refeq{legendreODE} becomes
\beq \label{eq:ODELegTransf}
\epsilon^2 \, \frac{d^2 w}{dx^2} + \frac{a^2 -x^2 + \epsilon^2}{(1-x^2)^2} \, w = 0
\eeq
free from the first order derivative term. This wave equation contains besides the kinetic term $w''(x)$  also potentials: in the notation of \cite{BenderOrzag} the leading potential is $Q$ with:
\beq 
Q(x) = -\frac{a^2-x^2}{(1-x^2)^2} 
\eeq
and a subdominant potential $Q_2$ proportional to $\epsilon^2$
\beq \label{eq:PotSubdom} Q_2(x) = - \frac{1}{(1-x^2)^2} \,.
\eeq
To leading order in the JWKB-method this subdominant potential is dropped although we will also consider the effect of $Q_2$ later on.

First there is a momentum 
\beq
p(x) =\sqrt{-Q(x)}  = \frac{\sqrt{a^2 -x^2}}{1-x^2} \,.
\eeq
Second there is a turning point when this momentum vanishes, i.e. at $x=\pm a$.  We only consider $0<x_0<x<a$.

\section{Semiclassical scattering quantization}
\label{sec:ScatPhase}
In the previous section we discussed the ray-limit of the symmetry reduced problem of a spherical cap in terms the asymptotics of the associated Legendre polynomials. In this section, this information is used  to derive the corresponding  full spectrum with a ray-limit involving geodesic polygons. First, we  state a simple asymptotic condition for an eigenmode in a form similar to that of  Bohr-Sommerfeld quantization in \sref{sec:BohrSomm}. Then we reformulate the resonance condition as one arising from scattering  in \sref{sec:ScatQuant} and check this condition numerically in \sref{sec:ScatNumTest}. We next progress from this condition for individual eigenfrequencies belonging to a given irreducible representation to the full spectral density in \sref{sect:DOSscalar}. We find, that the derived asymptotic density of states diverges in the limit of a hemisphere in \sref{sec:Divergence}.

\subsection{Resonance condition in the ray limit}
\label{sec:BohrSomm}
Following steps similar to the usual for  Bohr-Sommerfeld quantization \cite{BenderOrzag}  discussed in detail in \ref{app:JWKB},  the Dirichlet condition $\Psi = 0$ gives the following JWKB-condition for an eigenmode:
\beq \label{eq:ResoCondSimple}
\frac{1}{\epsilon} I_0 + {\pi \over 4}  = n \pi \,
\eeq
for $n \in \mathbb{Z}$ and  an {\it action} integral $I_0$ given as
\beq
\label{eq:LeadingAction}
\fl I_0 = \int_{x_0}^a \, p(t) \, dt= \int_{x_0}^a \frac{\sqrt{a^2 -t^2}}{1-t^2} \, dt = \acos\left({x_0 \over a }\right) -\sqrt{1-a^2} \, \atan\left( \frac{\sqrt{a^2 -{x_0}^2} }{\sqrt{1-a^2}\, x_0 }\right) \,.
\eeq
This condition follows from finding the roots of  the asymptotic form of the wave function
\beq
u \sim Y_{l m}(\theta, \phi=0) \approx \frac{1}{\pi} (a^2 -x^2)^{-1/4} \, \cos \left({I_0 \over \epsilon} -{\pi \over 4} \right)  \,
\eeq
taken  proportional to a spherical harmonic. We derive this leading asymptotic form   in  \ref{app:JWKB} as well as perturbative corrections for use later on in \sref{sec:JWKBCorr}.

\subsection{Scattering formulation}
\label{sec:ScatQuant}
We aim at deriving the spectral density for the spherical cap with the
spectrum given by the condition \refeq{eq:ResoCondSimple} using ideas from 
scattering  theory \cite{smilanskyCourse}.  

What is the scattering approach to quantization? This method exploits
the connection between an exterior scattering problem and an
associated interior resonator problem.  Thus, if we force an obstacle
from the exterior at an interior eigenfrequency the scattered wave
experiences no phase shift; we shall see this explicitly in our case.
The method is often referred to as \hbox{inside-outside} duality. We
will apply this method to find the spectral density for the interior
cap vibration problem using information from corresponding scattering
data.  We proceed to our particular geometry. There, formal scattering
for curved billiards on e.g. the sphere has already been discussed in
\cite{gutkin}.  For concreteness, we discuss the explicit form of the
scattering states in detail in \ref{app:ScatState}.

Thus in terms of scattering theory, we shall say that a condition for an eigenmode is that
the  {\it phase} $\Theta_m$ :
\beq \label{eq:SingleScatPhase} 
\Theta_m \equiv
\frac{2}{\epsilon} \,I_0 + {\pi \over 2} 
\eeq 
has to equal zero modulo $2 \pi$
\beq
\label{eq:ScatPhaseCond}
\Theta_m = 0 \qquad (\mbox{mod} \, 2 \pi) \,.
\eeq
First,  \refeq{eq:ScatPhaseCond} is formally equivalent to
 \refeq{eq:ResoCondSimple}; second, the discussion in
\ref{app:ScatState} shows that there {\it is} an underlying set of
scattering states with precisely the scattering phase $\Theta_m$.  
So we define a scattering matrix diagonal in the azimuthal quantum number
$m$ 
\beq 
\label{defScatPhase} \scatmat = (S_m)_{m \in \mathbb{Z}}
\qquad \mathrm{and} \qquad S_m = \exp(-i \, \Theta_m) \,.  \eeq 
Then
an eigenmode occurs when the corresponding scattering problem is
transparent \cite{smilanskyCourse}. The choice of sign in
\refeq{defScatPhase} agrees with treatments of the flat case, i.e. the
disc \cite{smilanskyCourse,smilanskyIddo}.

\subsection{Numerical test}
\label{sec:ScatNumTest}
As an example, there is an eigenmode for
  $l = 99.6428945787050$ and $m = 70$ to good approximation for $\theta_0 = \pi/3$, i.e.  $P_l^m(\cos \theta_0) \approx 0$. For these values of $l$ and $m$ we can find $\epsilon$ and $a$ from \refeq{eq:Epsilon} and \refeq{eq:AorPsi}. The corresponding semiclassical phase evaluates to $\Theta_m/(2 \pi) = 4.0014$, close to an integer.

\subsection{Spectral density}
\label{sect:DOSscalar}
The distribution function for the eigenfrequencies is captured by the density of states. For convenience, instead of frequencies the spectral parameter in this section is  the dimensionless wave number $\kappa=k R$.   For several spectral problems the density can be approximated by a decomposition  consisting of a smooth $\overline{\rho}$  and oscillating part $\tilde{\rho}$. The smooth part is the most studied \cite{SAFAROV} whereas less is known about the oscillatory part. For our system we turn the attention to the latter.

In the scattering formulation \cite{smilanskyCourse} the {\it fluctuating spectral density} becomes
\beq \label{eq:DOSGeneral}
\tilde{\rho}(\kappa) = -\frac{1}{\pi} \, \Im \sum_{n=1}^\infty \frac{1}{n} \, \frac{\partial}{\partial \kappa} \Tr \left( \scatmat^n \right) = \frac{1}{\pi} \, \Im \sum_{n=1}^\infty \frac{1}{n} \, \frac{\partial}{\partial \kappa} \Tr \left( \scatmat^{*n} \right) \,.
\eeq
The terms for the $n$'th power in \refeq{eq:DOSGeneral} correspond to orbits bouncing $n$ times as we shall see in the following. 

\subsubsection{Poisson summation and n'th trace}
\label{sect:traceN}
As the scattering matrix  \eref{defScatPhase} is diagonal the trace becomes 
\beq
\label{trSum}
\Tr \, \scatmat^{n }= \sum_m (S_m)^n = \sum_m \exp({-i n \Theta_m}) \,.
\eeq
In the general case  \refeq{trSum} is done  by Poisson summation
\beq \label{trSumPoisson}
\Tr \, \scatmat^{ n} = \sum_m (S_m)^n = \sum_M  \int_{-\infty}^{\infty} dm \,  \exp({i \, (-n \Theta_m- 2 \pi m M)})  \,
\eeq
and subsequently approximated by stationary phase.

\subsubsection{The stationary point corresponds to geodesic polygons }
\label{sec:SPA}

The saddle point condition for the trace \eref{trSumPoisson} becomes
\beq
-n \frac{\partial \Theta_m}{\partial m} = 2 \pi M \,.
\eeq
We calculate the derivative of \eref{eq:SingleScatPhase} using \eref{eq:LeadingAction}   
\beq \label{eq:Igrad}
 \frac{\partial I_0}{\partial m} = -\epsilon \, \atan\left( \frac{\sqrt{a^2 -x_0^2} }{\sqrt{1-a^2}\, x_0 }\right)=  -\epsilon \, \atan\left( \frac{\sqrt{\sin^2 \psi -\cos^2 \theta_0} }{\cos \psi \cos \theta_0 } \right)
\eeq
and find the condition
\beq
\frac{\pi M}{n} = \atan\left( \frac{\sqrt{\sin^2 \psi -\cos^2 \theta_0} }{\cos \psi \cos \theta_0 } \right) 
\eeq
or by 
elementary trigonometric manipulations 
\beq \label{eq:cosSemi}
\cos \psi = \frac{\sin \theta_0 \cos \frac{\pi M}{n}}{\sqrt{\cos^2 \theta_0 + \sin^2 \theta_0 \, \cos^2 \frac{\pi M}{n}}}
\eeq
and
\beq \label{eq:sinSemi}
\sin \psi = \frac{\cos \theta_0  }{\sqrt{\cos^2 \theta_0 + \sin^2 \theta_0 \, \cos^2 \frac{\pi M}{n}}} \,.
\eeq
These conditions are precisely those of {\it classical} geodesic polygons winding $M$ times around the north pole and hitting the boundary $n$ times, see \refeq{eq:cosClass}. For an example of a classical polygon orbit see \fref{pentagram} and \fref{fig:DosAndPoly} showing  the pentagram orbit.
\subsubsection{Action at stationary point}
\label{sec:LeadingAction}
We next discuss the constant term in the stationary phase approximation of \refeq{trSumPoisson}. 

The square of the denominator of  \refeq{eq:sinSemi} is rewritten  using \refeq{eq:E1DotE2}:  
\beq \fl
\cos^2 \theta_0 + \sin^2 \theta_0 \, \cos^2 \frac{\pi M}{n}=\cos^2 \theta_0  + \sin^2 \theta_0 \, {1+ \cos({2 \pi M \over n}) \over 2}  = \frac{1+\cos \chi}{2}=\cos^2 \frac{\chi}{2}
\eeq
with $\chi$ being the angle of a single geodesic segment \refeq{eq:SegmentAngle}. Then consider 
\beq \label{eq:I0Saddle}
I_0 =\acos\left( \frac{\cos \theta_0}{\sin \psi} \right) -{\pi M \over n} \, \cos \psi = {\chi \over 2} -{\pi M \over n} \, \cos \psi
\eeq
by \refeq{eq:LeadingAction}, \refeq{eq:sinSemi} and \refeq{eq:E1DotE2}. 
We proceed to calculate $-n \,\Theta_m-2 \pi m M$ in \refeq{trSumPoisson}  by \refeq{eq:SingleScatPhase}:  the   $2 \pi m  M$  cancels with the remaining term from  \refeq{eq:I0Saddle}  by \refeq{eq:AorPsi}.
Thus
\beq \label{eq:ActionAtSPA}
-n \Theta_m-2 \pi m M = -n {\chi \over \epsilon} - n {\pi \over 2}  \,.
\eeq
For the trace formula \refeq{eq:DOSGeneral}  eventually the conjugate scattering phase is used  corresponding to minus \refeq{eq:ActionAtSPA}. By  \refeq{eq:LengthSegment}, $n \chi/\epsilon = n \chi k R $ corresponds to the wave number times the total length of the geodesic polygon, i.e. the classical action of the geodesic polygon.  At this point we  identify the classical angular momentum $R p$ with multiples of the dimensionless wave mechanical, i.e. $\sqrt{l(l+1)}$, with $p R$ replaced by $k R$.

\subsubsection{Second variation}
From \refeq{eq:Igrad} the second derivative becomes
\beq \label{eq:Ihess}
\frac{\partial^2 I_0}{\partial m^2} =  \epsilon^2 \, \frac{x_0}{a^2 \sqrt{a^2-x_0^2}}= \epsilon^2 \, \frac{(\cos^2 \theta_0 + \sin^2 \theta_0 \, \cos^2 \frac{\pi M}{n})^{3/2}}{\cos^2 \theta_0 \sin \theta_0 \, \sin \frac{\pi M}{n}}
\eeq
when evaluated at the saddle point. 

We express this Hessian  using the classical anholonomy: 
\beq \label{eq:HessPhLag}
\frac{\partial^2 I_0}{\partial m^2} = - \frac{\epsilon}{2} \, \frac{\partial  \delta \phi}{\partial L_z} \,
\eeq
when comparing \refeq{eq:Ihess}  with \refeq{eq:PhaseLag}.  Such identifications are customary in the context of trace formulae for systems with continuous symmetries \cite{CreaghLittle,Creagh}.

As for the interpretation of the point of stationary phase in \sref{sec:LeadingAction}, we identify the classical angular momentum in \refeq{eq:HessPhLag}  with the dimensionless wave mechanical momentum $k R$. Alternatively, we could  have chosen  a dimensionless measure of orbit anholonomy with a particle of unit classical momentum: $\partial \delta \phi/\partial (\cos \psi)$, see \refeq{eq:PhaseLag}.

\subsubsection{Trace formula}
\label{sect:DOSscalarResult}
Collecting the previous results leads to the following oscillatory density of states in the variable $\kappa = k R$:
\beq
\label{eq:DOSOsc}
\tilde{\rho}(\kappa) = \sqrt{\frac{2}{\pi}} \, \sum_{M=1}^\infty \,\sum_{n=1}^\infty \, \sqrt{\left|\frac{\partial L_z}{\partial \Delta \phi}\right|} \, \chi \, (-1)^n\, \cos\left( n \chi \kappa - n {\pi \over 2} + {\pi \over 4} \right)\,.
\eeq
We summarize the various factors in this formula. In particular $\chi$ being the angle of a segment from the origin,  $  n \chi \kappa = k \cdot n R \chi $ is the classical phase proportional to the length of the geodesic polygon. The sign $(-1)^n$ is the phase shift from the Dirichlet boundary condition.
Furthermore the derivative ${\partial L_z}/{\partial \Delta \phi}$ is interpreted as the {\it accumulated} phase lag for a closed orbit with $n$ segments  $\Delta \phi \equiv n \,  \delta \phi $ using \refeq{eq:PhaseLag}.

The result \eref{eq:DOSOsc} for the density of states of the spherical cap
is consistent with the general $U(1)$-symmetry
reduced trace formulae discussed in
\cite{CreaghLittle,Creagh}. Similar agreements are found in the flat
case for a disc, both quantum and elastic \cite{smilanskyCourse,disc}.

\Eref{eq:DOSOsc} represents the result to leading order in $\kappa$. Later in \sref{sec:JWKBCorr}  we include the first perturbative correction.

The classical interpretation of  the spectral density also holds more generally:
much earlier in systems without symmetries Gutzwiller connected the fluctuations to classical periodic orbits \cite{gutbook} using the method of path integrals.

For a opening angles $\theta_0$ larger than $\pi/2$ one can always
associate the spectrum to that of a cap with opening angle
$\pi-\theta_0$ as
\beq
0=P_l^m(-x) =\pm P_l^m(x)
\eeq
with $x= \cos \theta_0$, already remarked in \cite{gutkin}. In
particular, the geodesic polygons of a small cap also governs the
spectral fluctuations of its complementary cap.

\begin{figure}
\begin{center}
\includegraphics[viewport=88 4 520 271, scale=0.7]{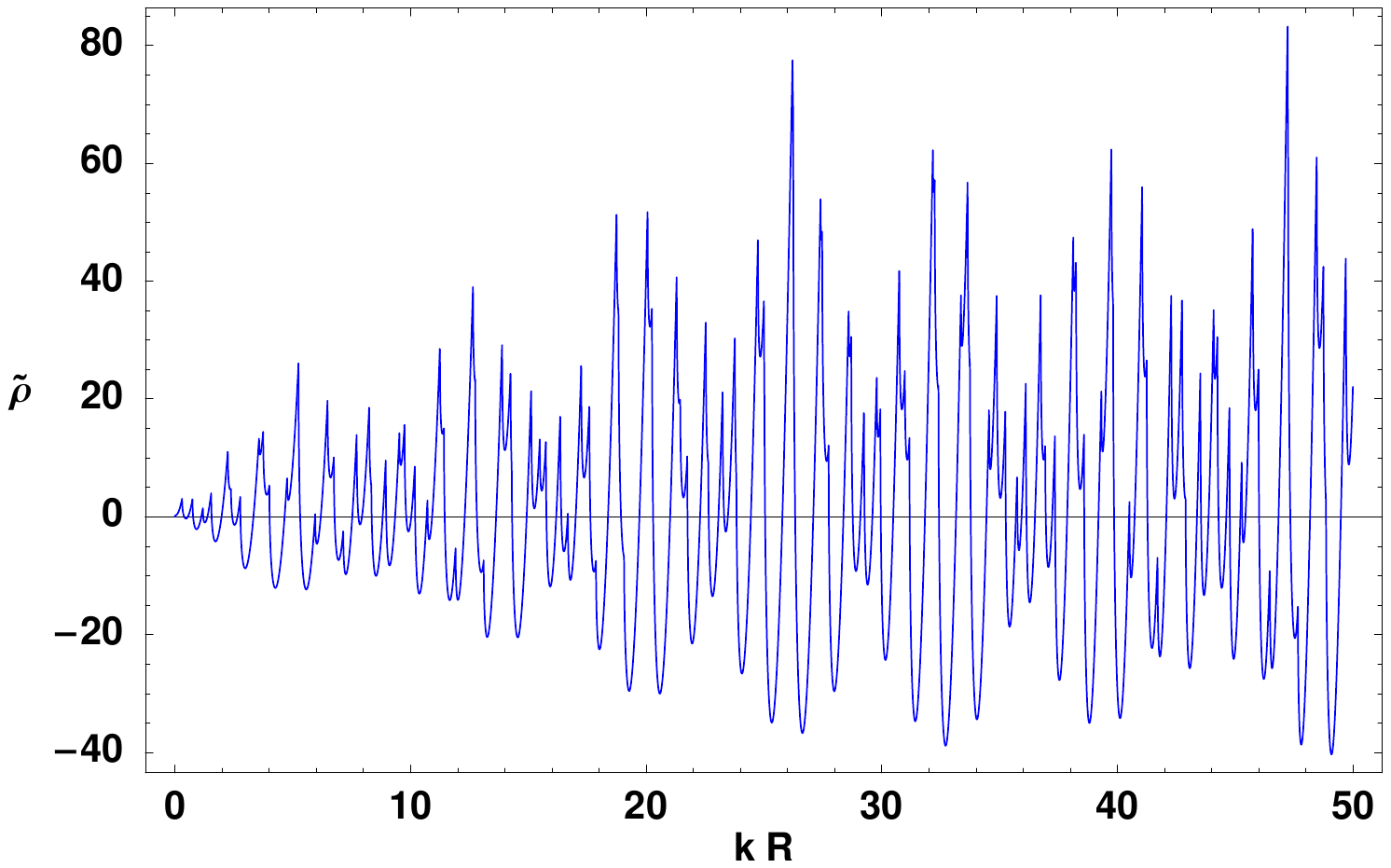}
\caption{\label{FdosGeneral} Spherical cap: Oscillatory density of states for opening angle $\theta_0 = \pi/3$}
\end{center}
\end{figure}

{\it Numerical checks}:
   By including a few orbits and many repeats it is possible to build up a distribution with sharp peaks, see \fref{FdosGeneral}.  This technique is discussed by e.g. \cite{brack} for the flat case of a disc.  We have performed this simple test as a check for both Dirichlet and Neumann.  In all cases at $\theta_0 = \pi/3$ the peaks fell approximately at the positions of the exact eigenfrequencies when including sufficiently many short orbits and repeats.

\subsection{Analysis of orbits as the opening angle changes}
\label{sec:Divergence}

\begin{figure}
\begin{center}
\includegraphics[viewport=88 4 520 340,scale=0.3]{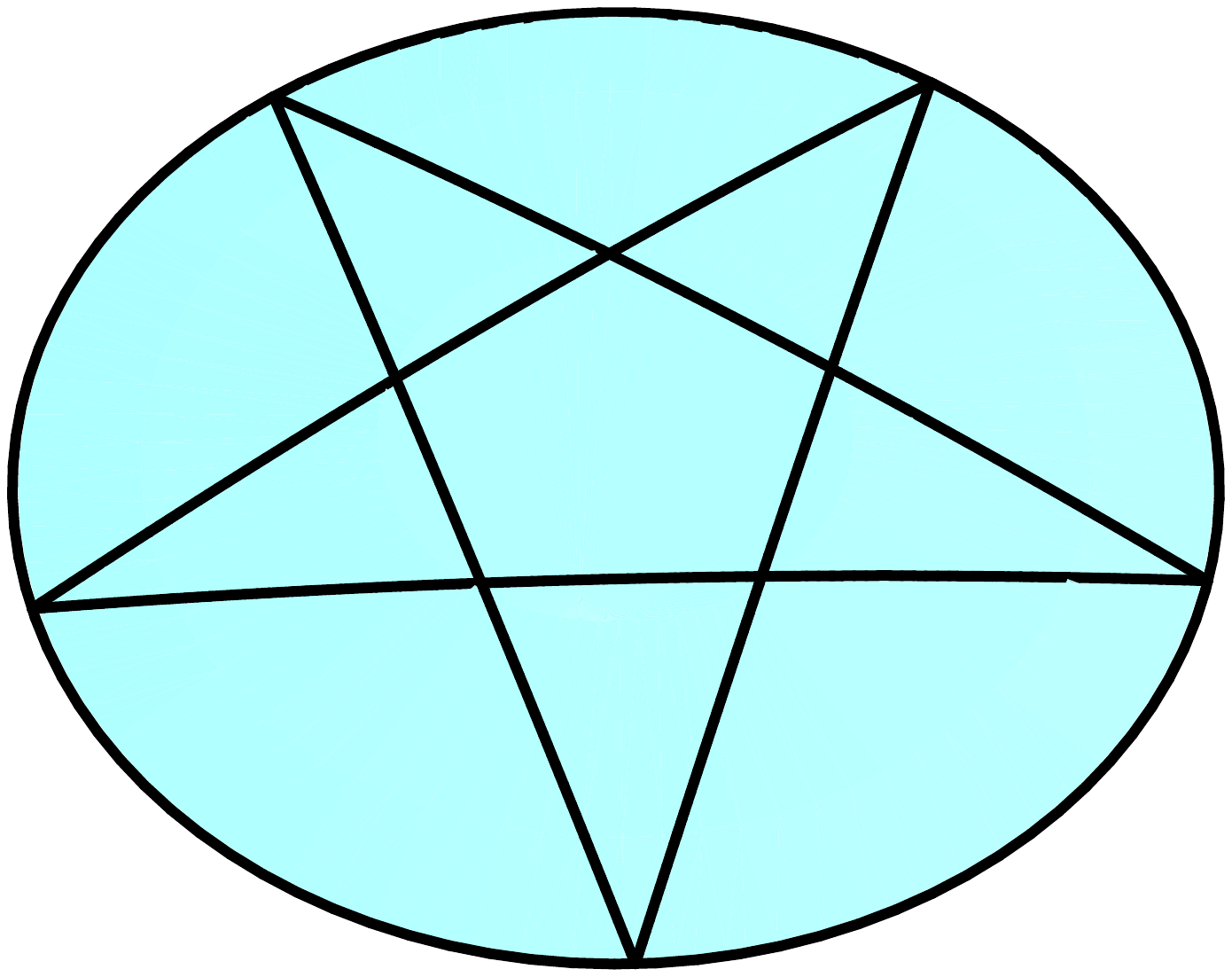} \includegraphics[viewport=88 4 520 340,scale=0.3]{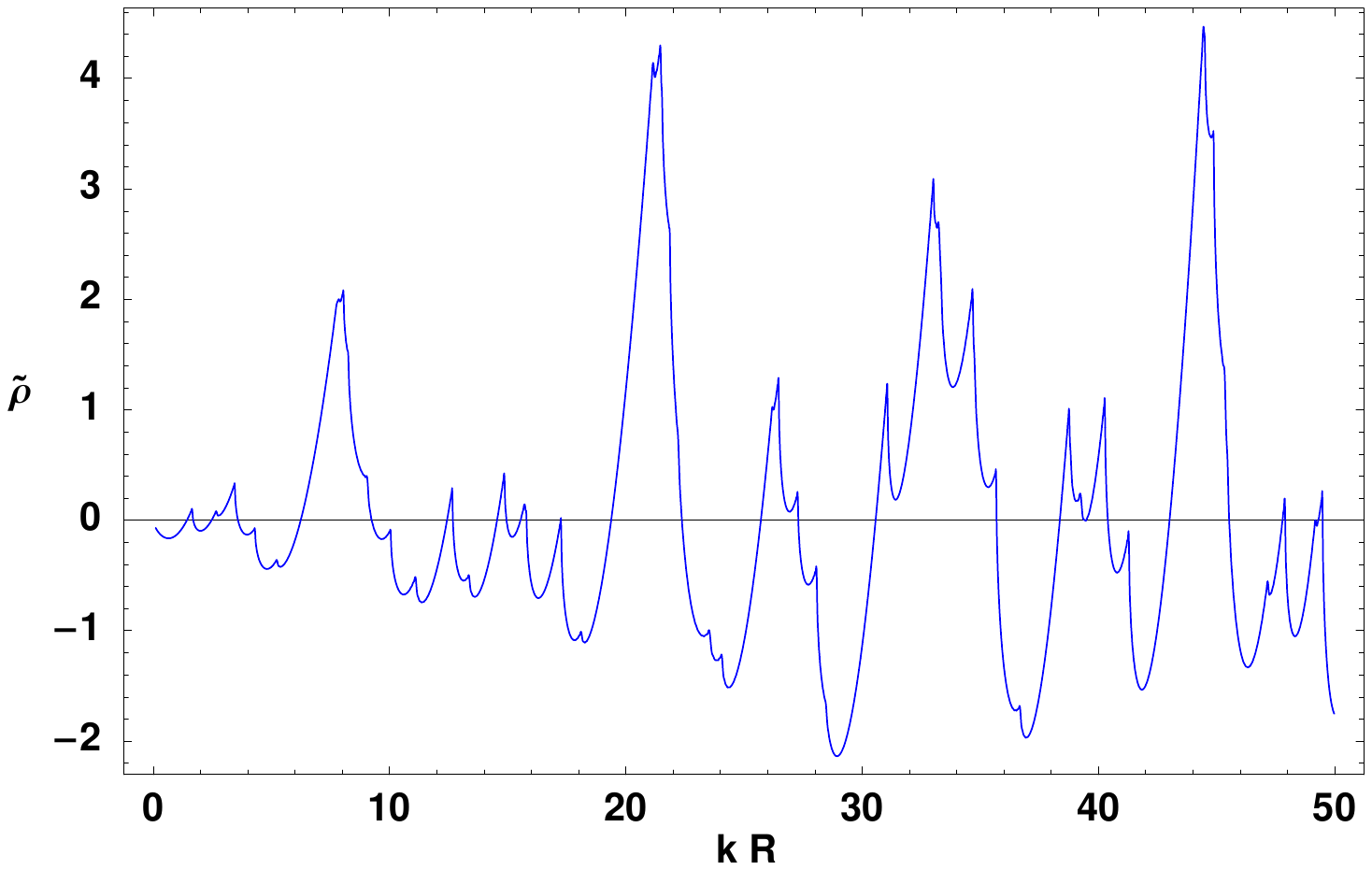}\\
\includegraphics[viewport=88 4 520 340,scale=0.3]{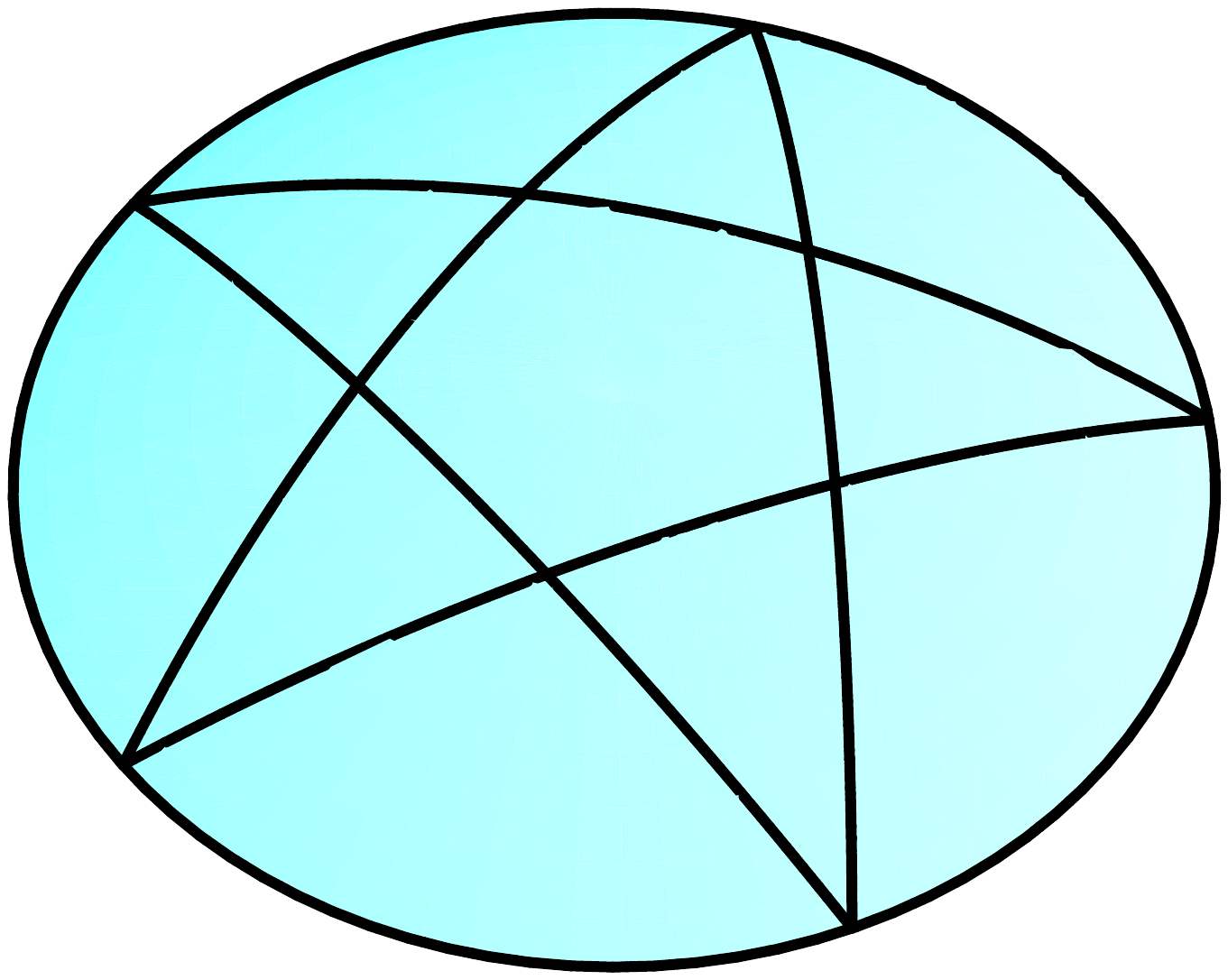} \includegraphics[viewport=88 4 520 340,scale=0.3]{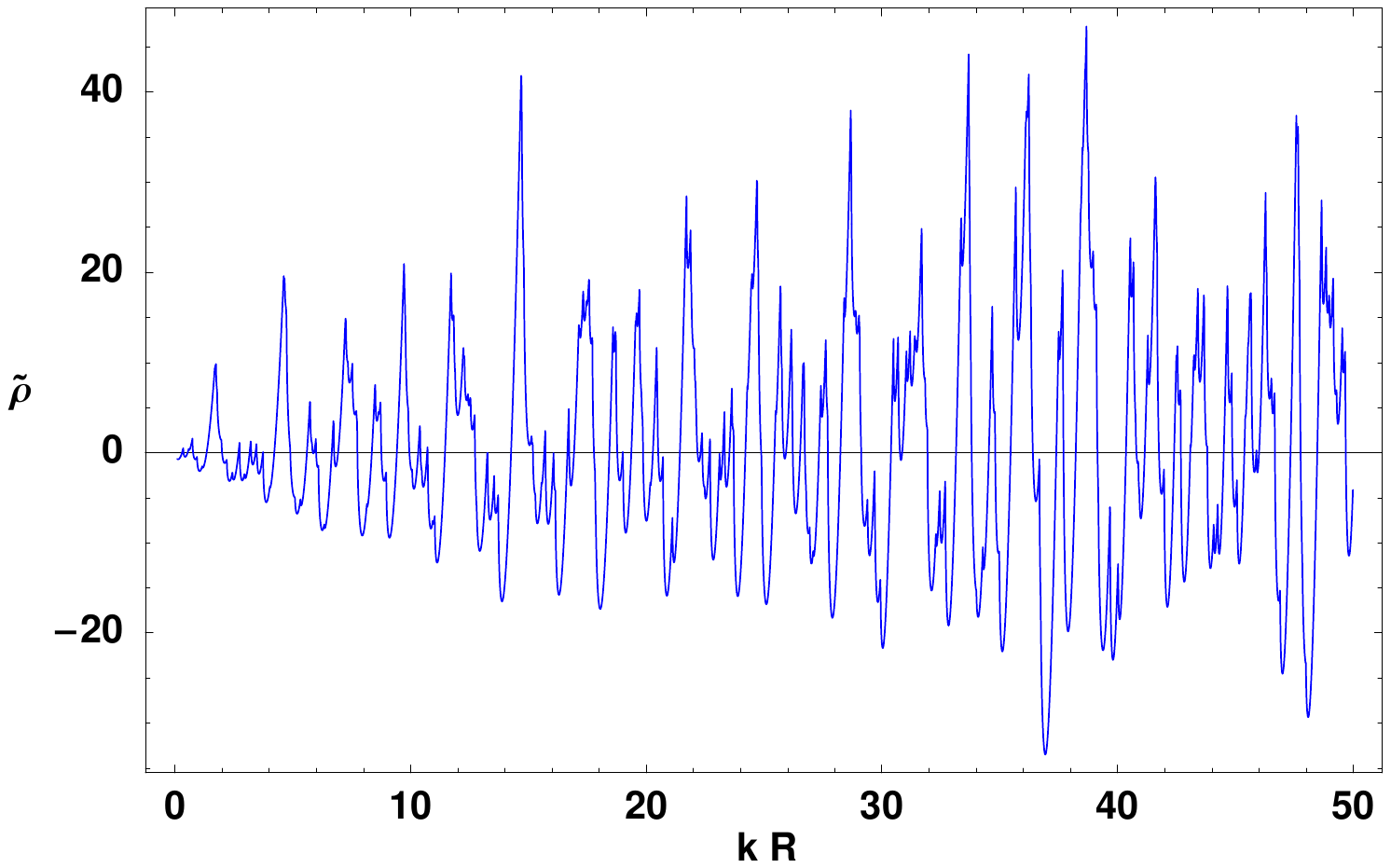}\\
\includegraphics[viewport=88 4 520 340,scale=0.3]{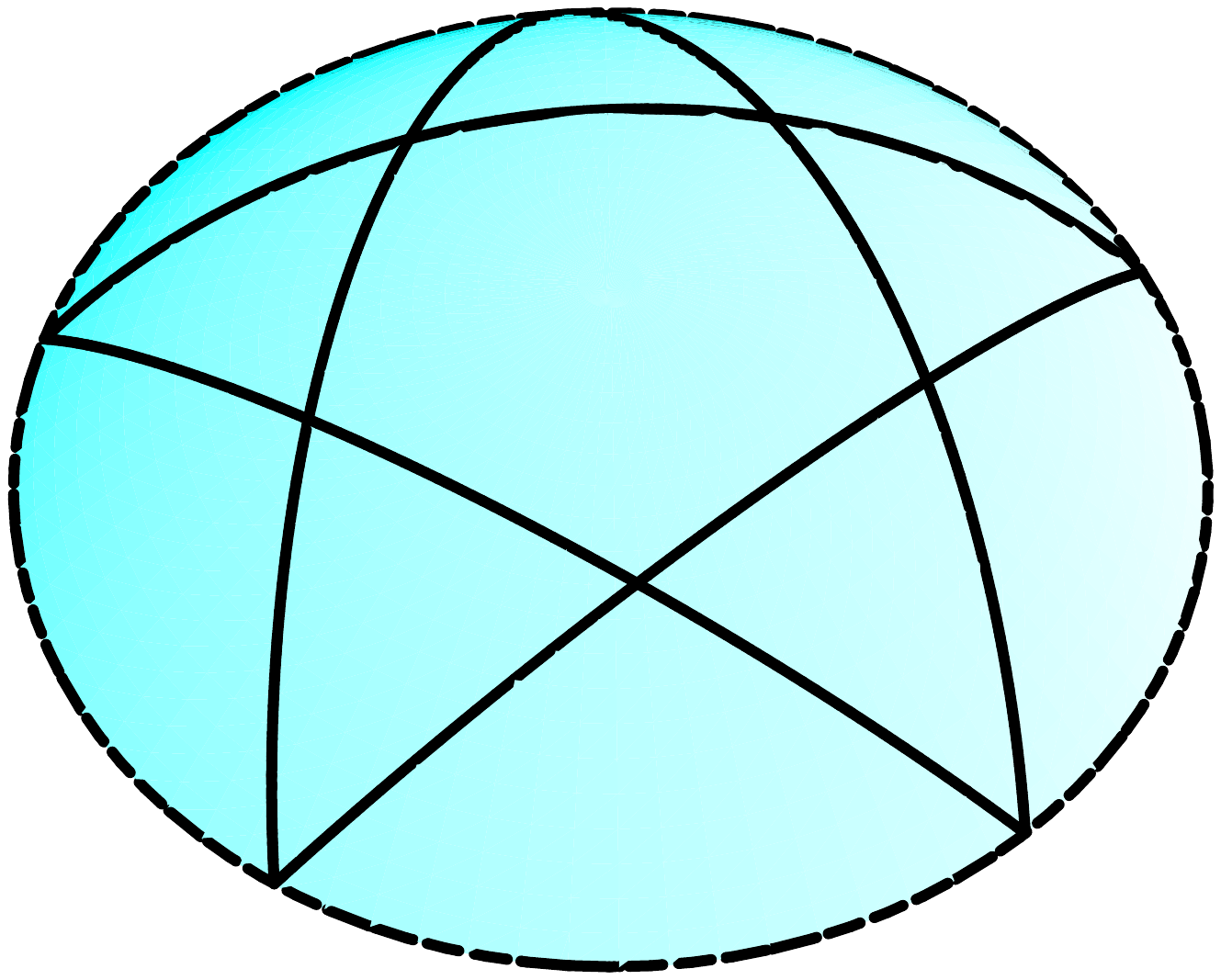} \includegraphics[viewport=88 4 520 340,scale=0.3]{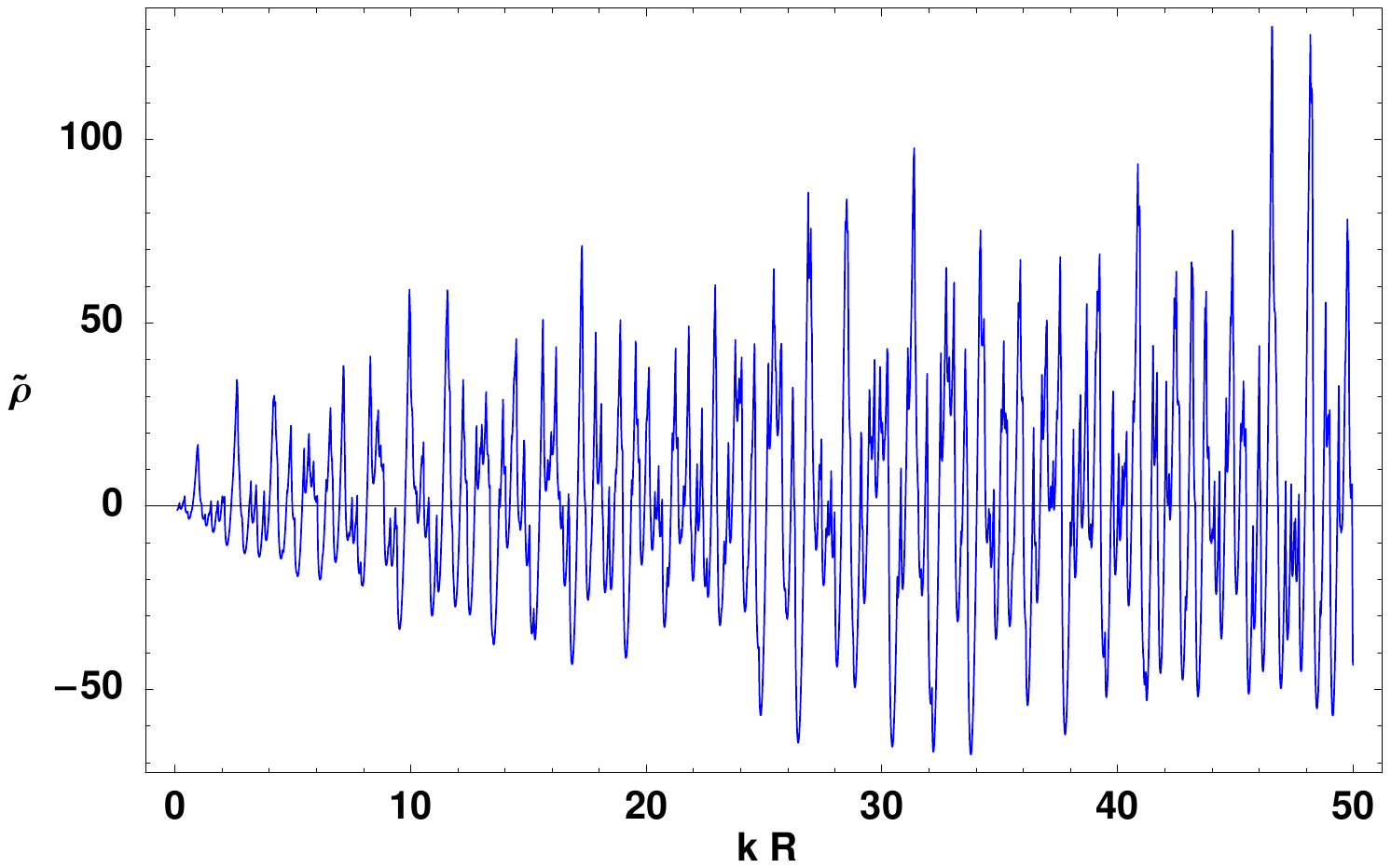}\\
\includegraphics[viewport=88 4 520 340,scale=0.3]{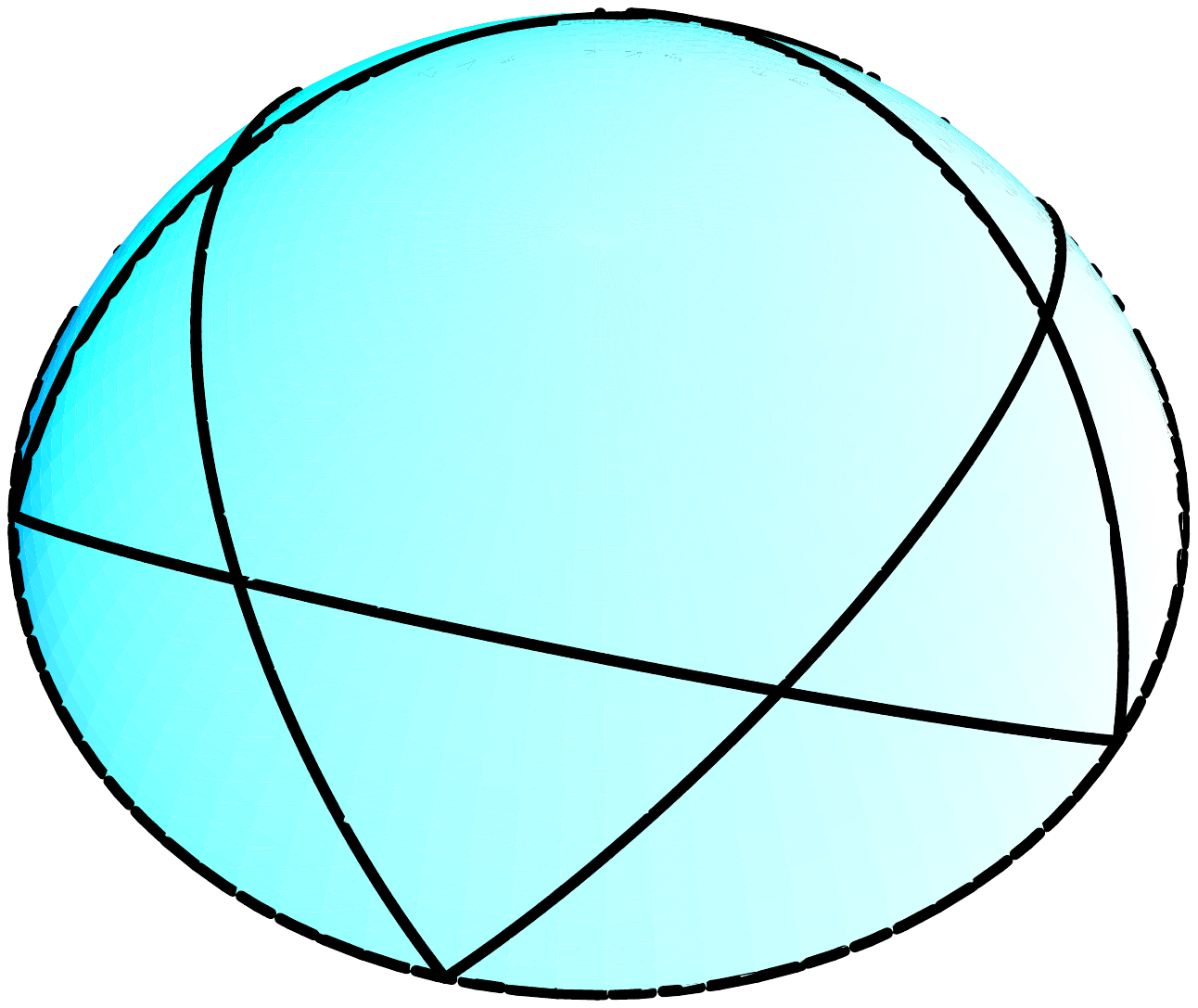} \includegraphics[viewport=88 4 520 340,scale=0.3]{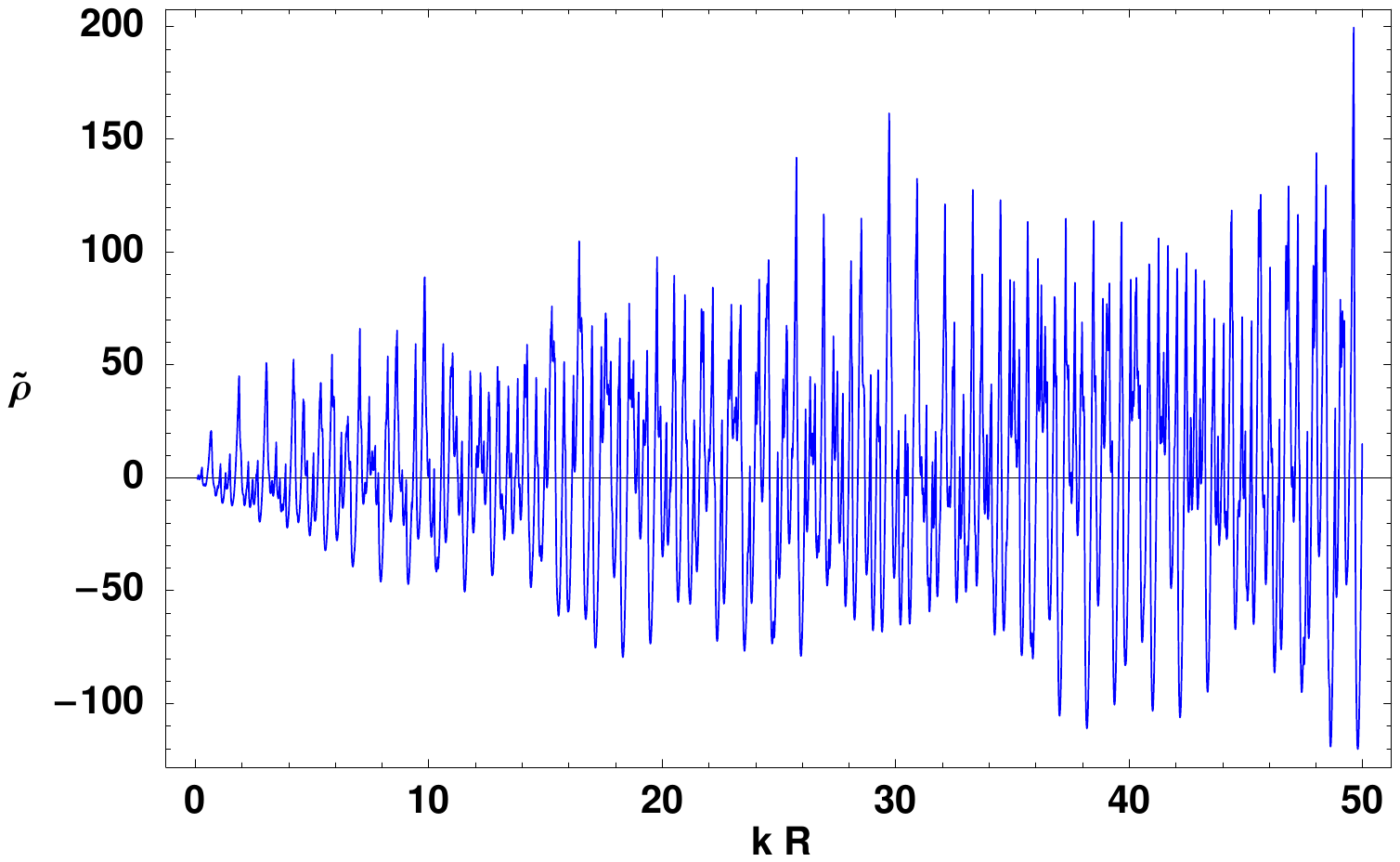}\\
\includegraphics[viewport=88 4 520 340,scale=0.3]{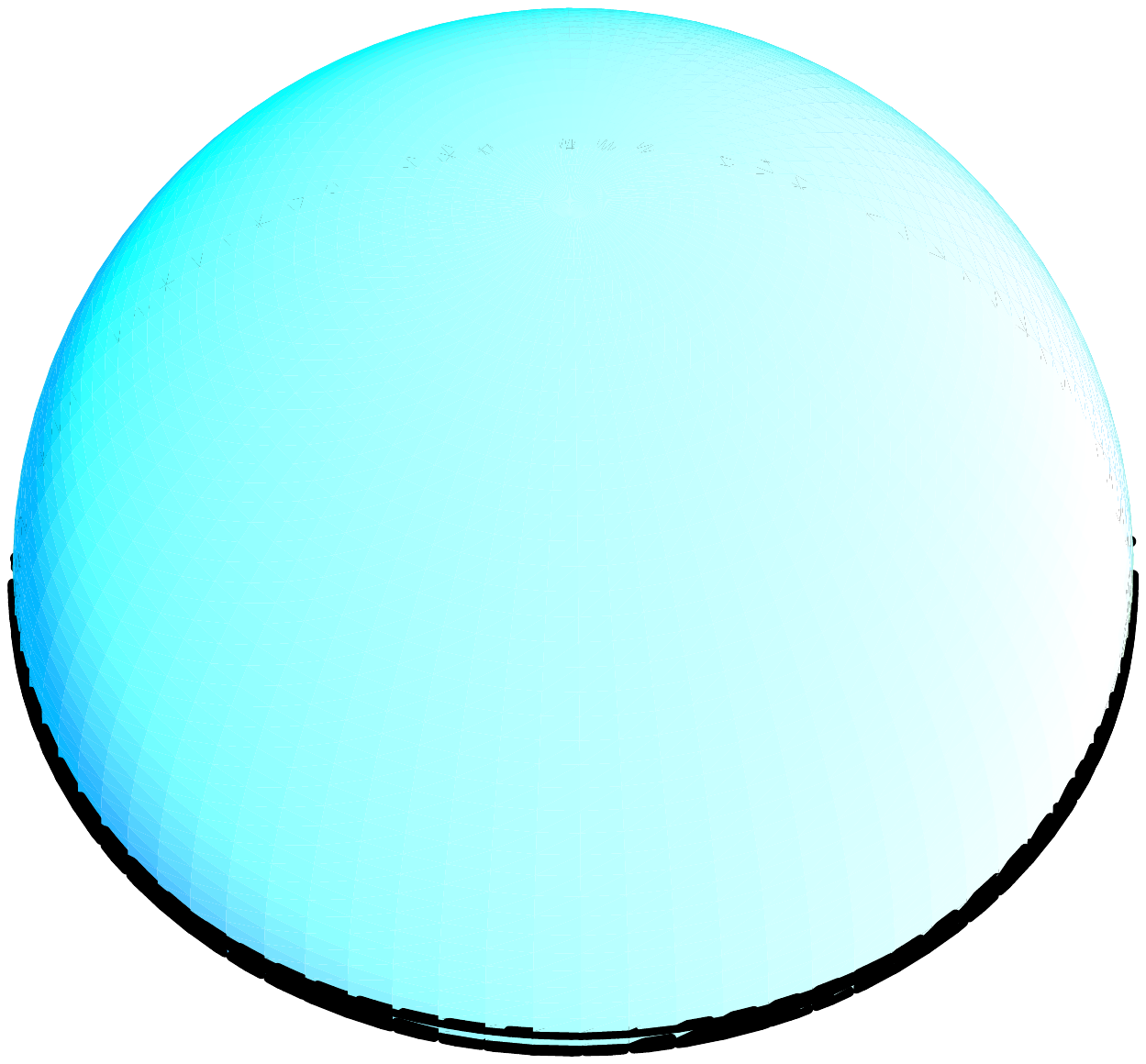} \includegraphics[viewport=88 4 520 340,scale=0.3]{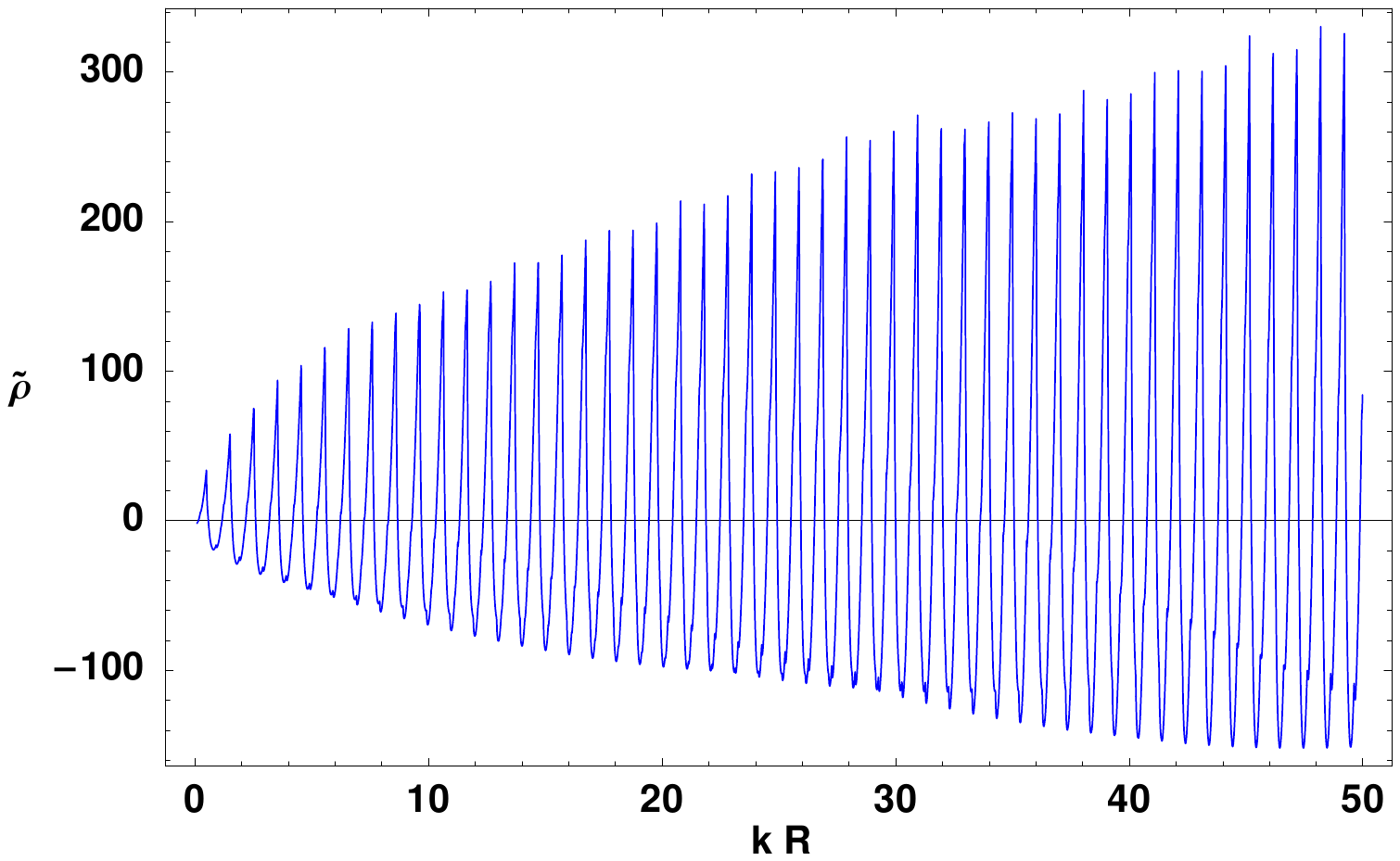}
\caption{\label{fig:DosAndPoly} The pentagram orbit and the density of states for fixed radius ($R=1m$) and varying opening angles $\theta_0 \in \{ 5.7^\circ, 26.5^\circ, 47.2^\circ, 68.0^\circ, 88.7^\circ \}$. }
\end{center}
\end{figure}

We now consider the trace formula \refeq{eq:DOSOsc} for general
opening angles and discuss what happens as the opening angle
changes. As the opening angle is varied the set of periodic orbits
changes accordingly. \Fref{fig:DosAndPoly} shows how the pentagram
orbit gradually converges from a pentagram in the plane towards an
orbit going twice around the equator. Likewise the semiclassical
weights associated to each orbit change, see \fref{fig:DosAndPoly}.
The calculations were done using the first 14 orbits having winding
numbers $(2,1), (3,1), \dots, (9,2),(9,4)$ and up to $80$ repeats.


Thus, the prefactor for each orbit family in the trace formula
is $(\partial L_z /\partial \Delta \phi)^{1/2} $. Using
\refeq{eq:PhaseLag} where $\delta \phi = \pi, \Delta \phi = 2 \pi$
being the angular increments for the diameter orbit: \beq (k
R)^{-1/2}\cdot \left( {\partial L_z  \over \partial \Delta
\phi} \right)^{1/2}_{\mathrm{diam}} = \frac{1}{2} \, d\theta^{-1/2} +
O(d\theta^{3/2}) \, .\eeq 
On the other hand non-diameter orbits have
weights going rapidly to zero as a function of the deficit in the
opening angle: \beq (k R)^{-1/2} \cdot \left({\partial L_z  \over \partial \Delta
\phi} \right)^{1/2}_{\mathrm{non-diam}} = O(d\theta) \,.  \eeq In conclusion,
in the limit of the hemispherical cap the diameter orbit controls the
spectral density. This will also show up in 
\sref{sec:Hemi}.

\section{The hemisphere}
\label{sec:Hemi}
In the hemispherical case, the spectrum can be found exactly: we  consider the spectrum in terms of  individual levels in \sref{sec:HemiPoint} respective in the form of a spectral density in \sref{sec:HemiDos}.

\subsection{Exact spectrum of hemisphere}
\label{sec:HemiPoint}
 For Dirichlet or
Neumann conditions the eigenfunction in the polar variable either is
odd or even. The parity of the Legendre polynomial is governed by \beq
P_l^m(-x) = (-1)^{l-m} \, P_l^m(x) \,.  \eeq Thus Dirichlet or Neumann
conditions hold when either \beq l-m = \mathrm{odd} \qquad \mathrm{or}
\qquad l-m = \mathrm{even} \,.\eeq with degeneracies $\nu_D=l$
respective $\nu_N=l+1$. Thus the spectra will display bunching of
eigenmodes at integer $l$ respective approximate half integer
$\kappa$.

\subsection{Hemisphere density of states}
\label{sec:HemiDos}
\begin{figure}[ht]
\begin{center}
\includegraphics[viewport=88 4 520 329,scale=0.7]{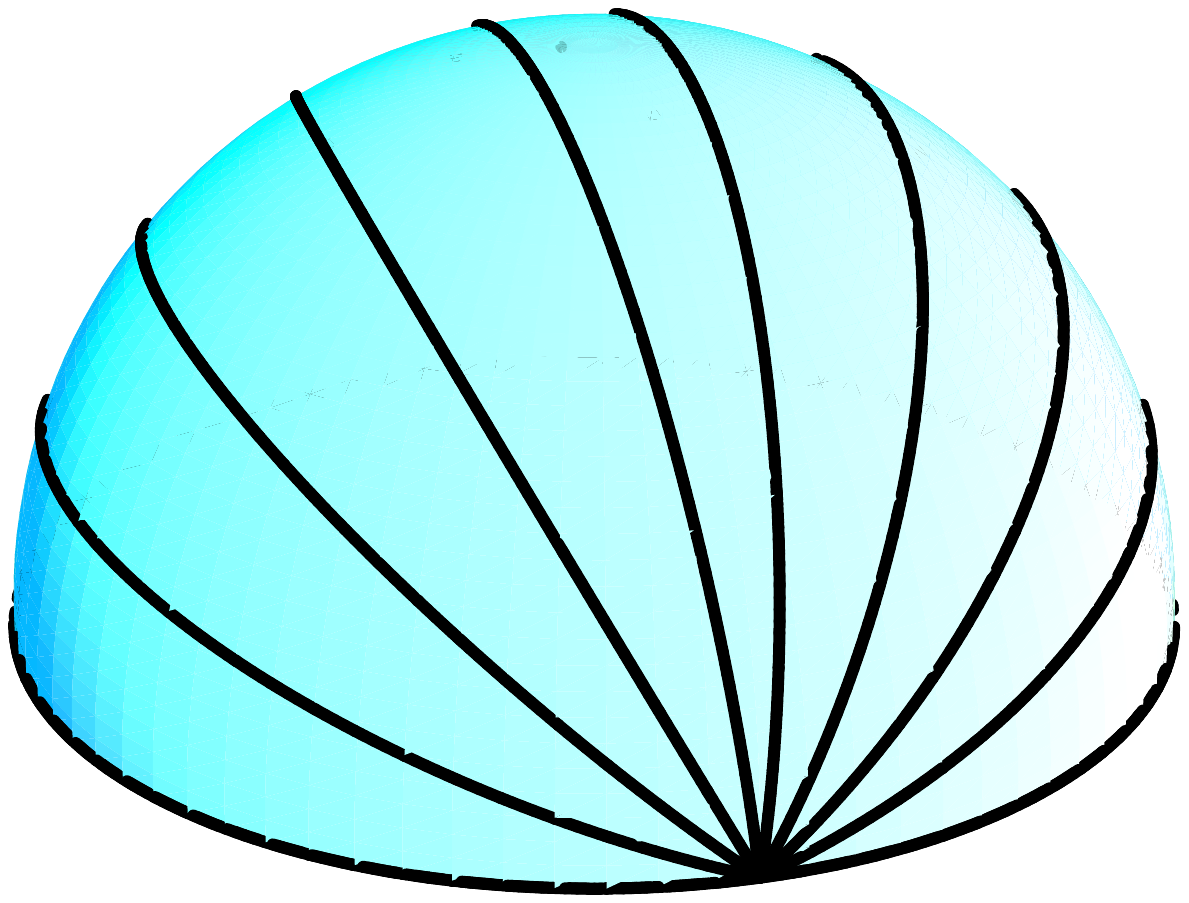}
\caption{\label{fig:multiRayHemi} The hemisphere: orbits with the length of the perimeter. }
\end{center}
\end{figure}
As we know the spectrum, the spectral density can  likewise be written exactly. Denoting the degeneracy  as $\nu$ the density becomes
\bea
\rho(l) &=& \sum_{n=0}^{\infty} \nu(l) \, \delta(l-n) \continue \nonumber
&=& \left\{ \begin{array}{ll} 
\sum_{n=0}^\infty l \, \delta(l-n) & \mbox{ (Dirichlet)} \\
& \\
\sum_{n=0}^\infty (l+1) \, \delta(l-n)  & \mbox{ (Neumann).}
\end{array} 
\right.
\eea
Poisson summation then gives the density in the Dirichlet case
\bea
\rho_D(l) &=& \sum_{N=-\infty}^{\infty} \, l \, e^{i 2 \pi N l} +\frac{l }{2} \delta(l) \continue \nonumber
&=& l + 2 l \sum_{N=1}^{\infty} \, \cos( 2 \pi N l)
\eea
respective in the Neumann case
\bea
\rho_N(l) &=& \sum_{N=-\infty}^{\infty} \, (l+1) \, e^{i 2 \pi N l} + \frac{l+1}{2}\, \delta(l) \continue \nonumber
&=& (l+1) + 2 (l+1) \sum_{N=1}^{\infty} \, \cos( 2 \pi N l) + \frac{1}{2} \, \delta(l) \,.
\eea
Switching to the spectral parameter $\kappa =k R = l+ {1}/{2} + O(l^{-1})$ for which $\kappa \, d\kappa = (l+1/2)\, dl$ holds exactly gives 
\beq
\label{eq:DosHemi}
\fl \rho_D(\kappa) \approx  \kappa-{1 \over 2} + 2 \, \left(\kappa-{1 \over 2} \right) \,  \sum_{N=1}^{\infty} \,  \cos \left( 2 \pi N \left(\kappa-{1 \over 2}\right)\right) 
\eeq
and 
\beq
\fl \rho_N(\kappa) \approx  \kappa+{1 \over 2} + 2 \, \left(\kappa+{1 \over 2}\right) \,  \sum_{N=1}^{\infty} \, \cos\left( 2 \pi N \left(\kappa+{1 \over 2}\right)\right) +\frac{1}{2}\, \delta \left(\kappa-{1 \over 2} \right) \,.
\eeq
The first two terms above  gives the smooth contribution to the density of states. Likewise the last terms correspond to the oscillatory part.

The smooth counting function $\overline{N}$ is the integrated density of states and becomes
\beq
\overline{N}_{D/N} = {k^2 R^2 \over 2} \mp {k R \over 2}   
\eeq
the first term being the available phase space volume divided by $(2 \pi)^2$ whereas the latter is the boundary correction well known from flat Helmholtz resonators $\mp {k}/{(4 \pi)} \, L$ with $L$ being the length \cite{SAFAROV,BaltHil}. Their sum corresponds to the spectral density of a full sphere. In particular, the sum of the boundary corrections  add up to zero corresponding to  no boundary.

The fluctuations are governed by the phase $2 \pi N \kappa =  k \cdot 2 \pi R N $ of the $N$'th repeat of an orbit with length of the circumference $2 \pi R$, see \fref{fig:multiRayHemi}.

\section{Comparing the two trace formulae}
\label{sec:CompareTrF}
In the previous \sref{sect:DOSscalarResult} and \sref{sec:HemiDos} we
presented two trace formulae for the oscillatory density of states for
a spherical cap. In this section we compare these two formulae.

{\it Orbits}:
The first trace formula 
\refeq{eq:DOSOsc} derived for a general cap opening angle contains 
a countable infinite number of orbits whereas the second \refeq{eq:DosHemi} only has
a single orbit with the length of the perimeter. 

{\it Weight versus wave number}:
In the hemisphere case this orbit has a weight proportional the spectral parameter and hence is of the order
of the smooth part of the density. However, in the non-hemisphere case families of orbits have weights only proportional to the square root of the spectral parameter. This is the conventional result for families of orbits in two-dimensional systems.

{\it Weight versus opening angle}:
Inspection of the first trace formula
\refeq{eq:DOSOsc} in \sref{sec:Divergence} revealed that non-diameter orbits are assigned
weights that vanish when the opening angle converges to $\pi/2$,
i.e. that of a hemispherical cap. On the other hand, the diameter
orbit has a weight which goes to infinity in this limit.  In
conclusion the first trace formula is a non-uniform asymptotic result
valid for $0 < \theta_0 < \pi/2$. On the other hand the second
trace formula is only valid at $\theta_0 = \pi/2$.
\begin{figure}[ht]
\begin{center}
\label{fig:CompareDOS}
\includegraphics[viewport=88 4 520 271, scale=0.5]{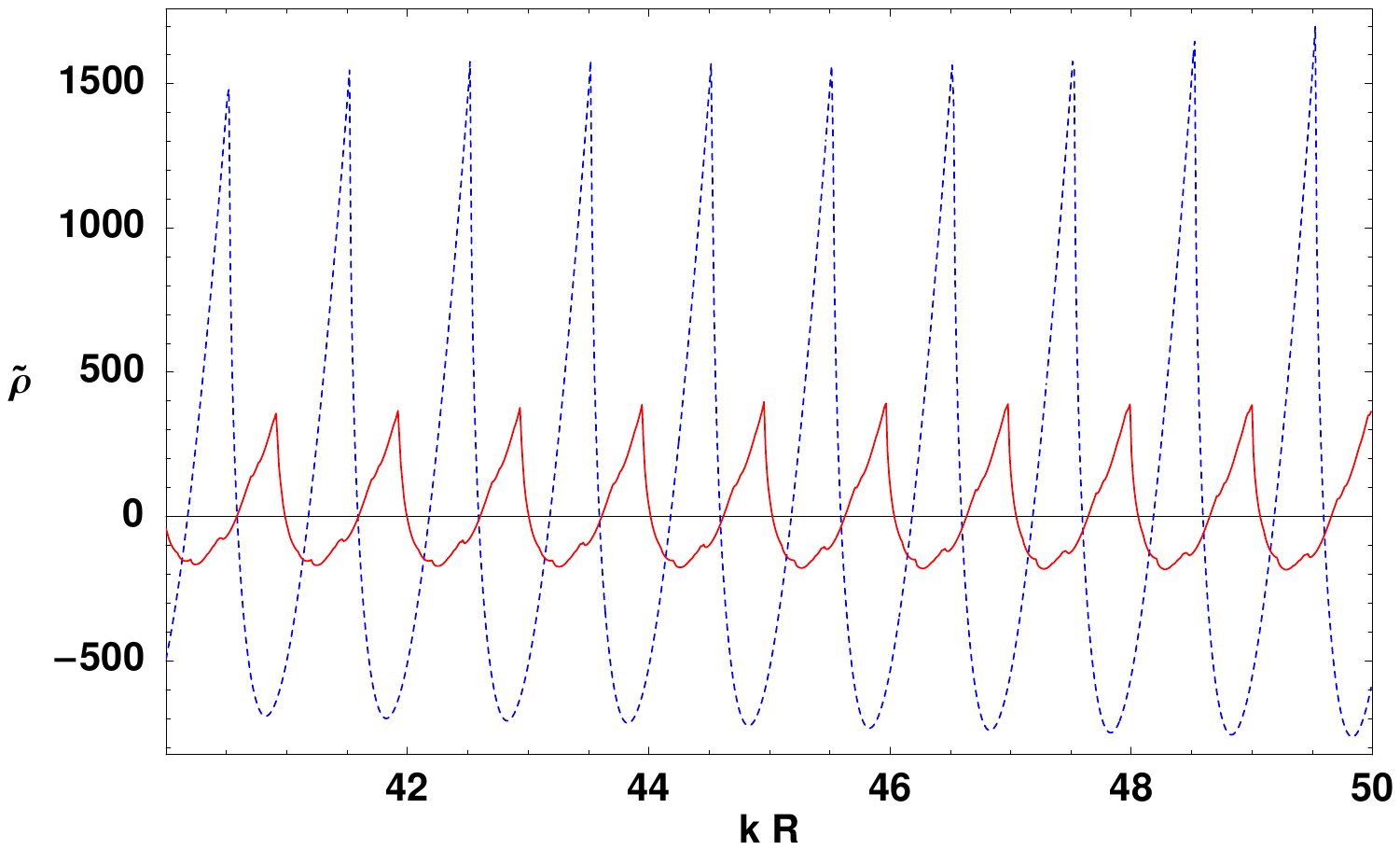}
\caption{Semiclassical density of states. Red/full line: $\theta_0=89.08^\circ$ and blue/dashed line: $\theta_0=89.95^\circ$.}
\end{center}
\end{figure}

{\it Near hemisphere}: 
When the opening angle approaches $\pi/2$  \refeq{eq:DOSOsc} has been checked to reproduce the exact peak positions, $\kappa \approx l+1/2 $ with $l\in \mathbb{N}$:  a little bit away e.g. $\theta_0=89.08^\circ$ is not sufficient whereas at  e.g. $\theta_0=89.95^\circ$ the predicted half integers occur, see \fref{fig:CompareDOS}.  This comparison was made  at relatively high wave numbers  with  $\kappa =k R \approx 45$.

\section{Leading correction}
\label{sec:JWKBCorr}
The previous calculations were all done to leading order in the
expansion parameter proportional to the inverse wave number $\epsilon
\equiv (k R)^{-1}$.  We shall now ask what is the  correction to
these results in this expansion parameter? There are two points of
interest: the individual sub-spectra for each $m$ (\sref{sec:PlmCorr}) and the total
spectrum on the level of the trace formula (\sref{sec:TrFormCorr}).

Besides the general question of how to calculate
JWKB-corrections, we comment on why that could be of interest: one
benefit of JWKB-corrections is that these may increase the validity of
trace formulae further down in the spectrum.  Ultimately
JWKB-corrections are believed to be particularly important in
dimensions larger than two \cite{smilanskyCourse}.  General results
for corrections are known in the case without boundaries
\cite{YCV73,Zel98} whereas only little is known with boundaries
(planar Dirichlet/Neumann case \cite{Zel04}).  By studying the
spherical cap we will get direct access to a simple situation for a
curved manifold with boundary.  In the present contribution we shall
obtain a non-trivial result already for Dirichlet conditions.

\subsection{Correction to leading asymptotics of associated Legendre polynomial}\label{sec:PlmCorr}
We are interested in the correction to the action $I$ when going from
the boundary $x_0$ to the turning point $a$ for the equation
\refeq{eq:ODELegTransf}. The discussion of JWKB-corrections when the orbit is
between two turning points \cite{BenderOrzag} is our starting point. We outline the corresponding theory below with details of the calculation  given in
\ref{app:JWKB}. 
However, for \refeq{eq:ODELegTransf} there is just a single
turning point and now also the subdominant potential $ Q_2(x)$ given
by \refeq{eq:PotSubdom} not discussed in \cite{BenderOrzag}.  The way
subdominant terms such as $Q_2$ enter is via the hierarchy of
transport equations that arises in the JWKB method.  To get the first
correction the second or higher transport equations must be
considered. In our case, $Q_2$ is present at the second transport
equation.

In the typical case {\it without} boundary several results are known
for subdominant corrections to the leading Hamiltonian.  For instance
\cite{ColDeVerd05} gives a systematic discussion of
$\hbar$-corrections considering a general series expansion, the {\it
Weyl symbol} of the corresponding Hamilton operator allowing for
momentum terms as well. For a general reference on JWKB-corrections see
\cite{froman}. In our case, however, we must also consider the
presence of a boundary $x_0$.

The result we find in  \ref{app:JWKB} is a second order correction of the action 
\beq I
\equiv I_0 + \epsilon^2 \, I_2 \,, \eeq where formally $I_2$ is found
by integrating $I_2'$ from the second transport equation.  Just as the
leading amplitude of the JWKB-function determined by the first
transport equation diverges near the turning point,  also the second
transport equation leads to divergences.  In our case the subdominant
potential $Q_2$ remains well behaved there.  
The singularities can be dealt with by for example transforming the problem to the Airy equation as in \cite{BenderOrzag} and is given in detail in \ref{app:JWKB}.  

To state the final result for the phase $I_2$ we shall need the expansion of  the
potential $Q$ in the coordinate $x$ around the turning point $a$:
\beq
\label{eq:PotExp}
Q(x) \approx \alpha \, (x-a) + \beta \, (x-a)^2 \,.\eeq 
Thus in the
transition region $x\approx a$ there is an approximate solution having
the functional form of an Airy function which in the region $x<a$
takes the form of a sine function.  That region also possesses a
standard oscillatory JWKB solution
\beq w_{osc}(x) \sim 
\Im \left[C \exp({i \, I}) \right] \eeq 
with the action $I$
calculated up to a point before the turning point ($a-\mu$ with $\mu
\rightarrow 0+$) 
\beq I=I(x)= \int_x^{a-\mu} I'(t) dt \,.  \eeq
We next match this oscillatory solution to the regular
Airy-solution by adjusting the pre-factor $C$.    
This matching modifies the second
order phase to 
\beq \eqalign{
\label{eq:JWKBCorr}
I_2 &\equiv \lim_{\mu \rightarrow 0+} \left(-\frac{5 Q'(x_0)}{48
  (-Q(x_0))^{3/2}}+\frac{1}{48} \int_{x_0}^{a-\mu}
  \frac{Q''(t)}{(-Q(t))^{3/2}} \, dt \right. \\ &\left. -\frac{\beta
  \, \alpha^{-3/2}}{12 \sqrt{\mu}}-\frac{1}{2} \int_{x_0}^a \,
  \frac{Q_2(t)}{(-Q(t))^{1/2}} \, dt \right).}  \eeq 
  The first three
  terms are stated in \cite{BenderOrzag} whereas the final term is the
  correction from the subdominant potential $Q_2$. 
  In \refeq{eq:JWKBCorr}
  the first two terms and the last correspond to the result obtained
  by formally integrating the second transport equation for $I_2$ and
  doing a partial integration, see \ref{app:JWKB} . In this context,
  the second term diverges as $Q(x)$ has a zero at the turning
  point. Given the expansion of $Q(x)$ by \refeq{eq:PotExp}, the third
  term coming from the matching exactly cancels
  this divergence and can be thought of as a regularization.

Using computer algebra and further manipulations \refeq{eq:JWKBCorr}
reduces in the present case to \beq \label{eq:I2Short} I_2=\frac{x_0
\left(3 a^4+2 \left(x_0^2-3\right) a^2+x_0^2\right)}{24 a^2
\left(a^2-x_0^2\right)^{3/2}} +\frac{\pi}{16} - \frac{1}{8} \, \atan
\left(\frac{x_0}{\sqrt{a^2-x_0^2}}\right) \,.  \eeq 
For an analytical check of the phase in the case of $x_0=0$ see \ref{app:JWKB}.

As a numerical example, we continue that of the end of
\sref{sec:ScatPhase} with  \hbox{$l = 99.6428945787050$, $m = 70$ and
$\theta_0 = \pi/3$}.  Including the correction $I_2$ for the action
leads to a scattering phase in units of $2 \pi$: $\Theta_m/(2 \pi)
=3.999993$. Further numerical checks are given in
\ref{app:ScatPhNumeric}.

\subsection{Effect of correction on the  the trace formula}
\label{sec:TrFormCorr}
The effect of the subdominant potential is to alter the saddle point position $m_0$ to
\beq
m^* = m_0 + \epsilon^2 m_2
\eeq
to second order but this only  affects the phases to fourth order:
put
\beq
\label{eq:DefS}
S = 2 n (I_0+\epsilon^2 I_2) \equiv S_0 + \epsilon^2 S_2   \,.
\eeq 
Then as the original
phase function is stationary \beq S_0^* = S_0(m_0) + (\partial_m
S_0(m_0)) \epsilon^2 \, m_2 + O(\epsilon^4)= S_0^* + O(\epsilon^4) \,
, \eeq whereas the new part leads to \beq S_2^* = \epsilon^2 S_2(m_0)
\,.  \eeq Finally, there is a contribution coming from a saddle point
correction of the original integral, i.e. by expanding to fourth order
in $m$ around $m_0$, $S_0^{(4)}$. There also other fourth order
moments. However, these are of lower order in $\epsilon$ and can be
neglected.

At the point of stationary phase
\bea 
\label{eq:CountwCorr} \fl
\int dm \, \exp({-i n \Theta_m}) &=& \int dm \, \exp\left({-i {S \over \epsilon}}\right) \continue
& &  \fl  \approx \left(  \frac{2 \pi \epsilon}{|S''_0|}  \right)^{1/2} \, \exp\left({i n {\pi \over 2} -i {\pi \over 4}}\right) \, \exp\left({- i \left({S_0 \over \epsilon} +\epsilon S_2 \right)}\right)   \, \exp\left({\sqrt{i} \epsilon   \,\frac{ S_0^{(4)}}{8 S_0^{(2) 2}}}\right) \,.
\eea
For a given orbit family, the factor $S_2$ only alters the phase whereas the latter factor also alters both   modulus and phase.

The density of states is calculated using \refeq{eq:DOSGeneral}: essentially a derivative with respect to $\kappa = 1/\epsilon$ is performed on \refeq{eq:CountwCorr}. Furthermore the derivatives of the action coefficients $S_i$ are  pertaining to $m$ and for these 
\beq
\frac{\partial}{\partial m} = -\epsilon^2 \,  \frac{m}{a} \, \, \frac{\partial}{\partial a} \,.
\eeq
In particular $S_0'' \sim \epsilon^2$. Thus apart from factors
\beq
\Tr \, \scatmat^n \sim \epsilon^{-1/2} \, \exp\left(-i \left({S_0 \over \epsilon}+S_2 \epsilon \right) +\xi \epsilon \right)
\eeq
with 
\beq
\xi = \alpha_1+i \alpha_2 \qquad \mathrm{and} \qquad \alpha_1=\alpha_2 = \frac{1}{\sqrt{2} \,8 } \, \frac{S_0^{(4)}}{S_0^{(2) 2}} \,.
\eeq
Hence the perturbative correction to the density of states goes as
\beq
\widetilde{\rho}(\kappa) \sim \partial_\kappa (\Tr \, \scatmat^{*n} ) \sim \sqrt{\kappa} \, S_0 \exp\left({\alpha_1 \over \kappa} \right) \, \cos \left( \kappa S_0 + {\Delta S \over \kappa} \right) \,,
\eeq
where 
\beq
\Delta S = S_2-\alpha_1-\frac{1}{2 S_0} \,.
\eeq
Incorporating all pre-factors and constants yields
\beq \fl
\widetilde{\rho}(\kappa) =  \sqrt{\frac{2}{\pi}} \, \sum_{M=1}^\infty \,\sum_{n=1}^\infty \, \sqrt{\left|\frac{\partial L_z}{\partial \Delta \phi}\right|} \, \chi \, (-1)^n\, \exp\left({\alpha_1 \over \kappa} \right) \, \cos\left( n \chi \kappa - n {\pi \over 2} + {\pi \over 4} + {\Delta S \over \kappa} \right)\,.
\eeq
From \refeq{eq:DefS} we notice the dependence on the number of repeats in the $S_i \sim
n$. Therefore $\Delta S = S_2 + O(n^{-1})$, so $S_2 = 2 n I_2$ with
$I_2$ given by \refeq{eq:I2Short} in the case of the spherical cap
remains the most important for high repeats.

\section{Summary and discussion}
\label{sec:conclusion}
We derived a trace formula for the spectral fluctuations of a
spherical cap of general opening angle. The
fluctuations in the spectral density were found to be governed by
geodesic polygons.   The method used is based on scattering referred to as
inside-outside duality with a discussion of the explicit scattering
states. Next the exact case of a hemisphere was introduced and there
the spectral fluctuations were found to be governed by a single orbit of
stronger weight than in the non-hemisphere case. The leading
result for general opening angles was compared to the hemispherical
case at relatively high wave numbers $k R \approx 45$. Although
non-uniform in nature, the general  trace formula without
JWKB-correction in the limit of opening angles close to $\pi/2$ also
exhibits peaks at positions corresponding to the hemispherical cap's
density of states. Finally, the leading JWKB-correction was discussed and
incorporated.

We used inside-outside duality in the derivation
of the trace formula.  The scattering states needed for this are
perhaps not so familiar to physicists but these functions have already been
introduced in seismology  \cite{DT98}. Furthermore
inside-outside duality for billiards in curved spaces has been
discussed recently in the thesis \cite{gutkin}. Despite that
inside-outside duality is just one possible method for this problem of
a spherical cap, it should also be mentioned that the method allows
for generalizations and clear interpretations in more complicated
cases such as systems of partial differential equations.  

Thus, although the leading result for the spectral density in
principle can be obtained by a general theory of symmetry reduced
trace formulae by \cite{CreaghLittle,Creagh} it is not always obvious
how to generalize these results in applications for wave equations
different from the Schr\"{o}dinger equation: which reflection
coefficients to use and what is the proper concept of the anholonomy
entering for the overall amplitude of a family of orbits in the
presence of for example ray-splitting.  By construction, the
scattering formalism automatically yields unitary reflection
coefficients related to for example probability or energy flux.
Furthermore, for the present work we have used the scattering method as
a vehicle to go beyond the leading results and incorporate
JWKB-corrections. In particular, such corrections to the spectral
density cannot at present be obtained from the works of
\cite{CreaghLittle,Creagh} which pertain to the leading result only.
We should also mention, that the scattering method can be extended to
general shapes without symmetries by considering the scattering of
suitable exterior states by for example attaching  wave-guiding leads
\cite{smilanskyCourse}.

We expect the scattering method on spheres as discussed in this
article and \cite{gutkin} could be generalized to the case of a sphere
with multiple circular holes. Here, addition formulae for the Legendre
polynomials for non-integer angular momentum $l$ \cite{MO53} would
lead to multiple-scattering expansions similar to those in flat space
between discs or spheres \cite{W99}. 

As mentioned the work in this article has partly been motivated
by Ellegaard's experiments on elastic shell caps. His group has
studied plates and three-dimensional elastic resonators see
\cite{Ell95,Ell96,ESB01} but is at the present time of writing
investigating shells as well. In this context, the work in this
article on the curved scalar Helmholtz equation and those in the flat
case \cite{disc,couch, HB} for two-dimensional elasticity show that
derivations of trace formulae in more general settings are
possible. Thus the combined case of curved elasticity \cite{kraus},
i.e. elastic shell caps, is indeed one generalization.

\ack
We thank the Swedish Research Council for financial support.
One of us (TG) also acknowledges support from the Deutsche
Forschungsgemeinschaft (Sonderforschungsbereich Transregio 12).

\appendix
\section{The JWKB solution}
\label{app:JWKB}

\subsection{Hierarchy of JWKB equations}
Using the oscillatory ansatz 
\beq 
\label{eq:AnsatzJWKB}
w = C \exp\left(i \left({I_0 \over \epsilon}+I_1 + I_2
\epsilon\right) \right) \eeq for the second order ordinary differential equation in the notation of
\cite{BenderOrzag} 
\beq 
\label{eq:ODEbasic}
\epsilon^2 w''= (Q+ Q_2 \epsilon^2)w \eeq
gives the eikonal equation \beq (I_0')^2+Q =0 \eeq and the transport
equation: \beq -2 I_1' I_0'+i I_0'' =0 \eeq respective the second
transport equation  
\beq 
\label{eq:2ndTransp}
-2 I_2' I_0' +i I_1'' - (I_1')^2 - Q_2=0 \,.  \eeq 
containing the correction from the subdominant
potential.
\subsection{Leading order}
We fix the action $I_0$ by integrating between $x$ and the turning point
$a$  with the result \eref{eq:LeadingAction}. Likewise
$I_1=i \log(-Q)^{1/4}$. From \refeq{eq:AnsatzJWKB} and \refeq{eq:NewWaveFct}, a real JWKB solution in the
oscillatory region away from the turning point is given up to two
unknown constants $A$ and $\delta$ as \beq \fl u = A \,
(1-x^2)^{1/2}\,(-Q)^{-1/4} \, \cos\left({I_0 \over \epsilon} +\delta\right)= A \,
(a^2-x^2)^{-1/4} \, \cos\left({I_0 \over \epsilon} +\delta \right)\eeq where $A$ is an
overall amplitude and $\delta$ is a phase shift.

One way of fixing $A$ and $\delta$ is by matching the wave function to
an exact solution, here taken as $\pi Y_l^m(\theta,\phi=0)$ with
\beq
Y_l^m(\theta,\phi=0) = (-1)^m \, \sqrt{\frac{2 l +1}{4 \pi}}   \sqrt{\frac{\Gamma(l-m+1)}{\Gamma(l+m+1)}} \,  P_l^m(\cos \theta) \,,
\eeq at the
point $x=0$ \cite{BPC96,MO53} using: \beq \label{eq:Leg1at0}
P_l^m(0)=\frac{2^m}{\pi^{1/2}} \,
\frac{\Gamma{(\frac{l+m+1}{2}})}{\Gamma{(\frac{l-m+2}{2}})} \,
\cos\left(\frac{\pi}{2} (l+m)\right) \eeq and \beq
{P_l^m}'(0)=\frac{2^{m+1}}{\pi^{1/2}} \,
\frac{\Gamma{(\frac{l+m+2}{2}})}{\Gamma{(\frac{l-m+1}{2}})} \,
\sin\left(\frac{\pi}{2}(l+m)\right) \,.  \eeq 
To proceed with the match, the gamma functions in the normalization of the spherical harmonics are expressed via the duplication formula  \cite{abramov}:
\beq
\label{eq:GammaDupl}
\Gamma(2 z) = \frac{1}{\sqrt{2 \pi}} \, 2^{2 z -1/2} \, \Gamma(z) \Gamma\left(z+{1 \over 2}\right) \eeq
and
subsequently
ratios of gamma functions are approximated
with \beq \frac{\Gamma(z+a)}{\Gamma(z+b)} \approx z^{a-b} \eeq valid
for large $z$, see \cite{abramov}. As in the main text $l$ and $m$ are assumed asymptotically large with a fixed ratio. 

Next, the leading semiclassical phase at
$x=0$ evaluates to 
\beq 
\label{eq:I0at0}
I_0 = \frac{\pi}{2} (1-\cos \psi) \eeq which
together with $1/\epsilon \approx l+1/2$ enters for the shift
$\delta$.  Finally $\cos \psi /\epsilon \equiv m$ to
all orders by definition \refeq{eq:AorPsi}.

Thus, for $A \equiv 1$
and $\delta \equiv -\pi/4$ the wave function matches \beq u \approx
\pi \, Y_{l m}(\theta, \phi=0) \,.\eeq 
In particular, the condition for an eigenmode is \beq {I_0 \over \epsilon}
-{\pi \over 4} = {\pi \over 2} + n \pi \eeq for $n \in \mathbb{Z}$.

\subsection{The first correction}
We  turn to the case where the sub-leading phase $I_2$ is included.
We distinguish between the oscillatory region ({\it osc}), the transition region ({\it trans}) and the decaying region ({\it exp}). We shall not be concerned with the decaying region discussed in detail in 
\cite{BenderOrzag}  (chapter 10.7).  Instead,  we focus on the oscillatory case from which it will be clear that  the subdominant potential $Q_2$ can be included in the discussion as well.

\subsubsection{Phase}
We  first remark, that we can always relate the solution at the boundary point $x_0$ with that at $x$ close to the turning point.
The  connection is by multiplication of the JWKB-factor 
\beq
w_{osc} (x_0) = \Im \left[ \exp\left(i \Delta I(x_0,x) \right) \, w_{osc}(x)  \right] \,.
\eeq
with
\beq
\Delta I (x,x_0)  \equiv  \int_{x_0}^x ds \, {dI \over ds} \,.
\eeq
Thus eventually all actions are continued further to the boundary as in \refeq{eq:JWKBCorr}.

Hence, the transport equations can be integrated formally, at least outside the transition region, whereas  these diverge at the turning point and require  discussion.  
 In detail, to get the action to first
order in $\epsilon$ we need to integrate 
\bea \label{eq:I2Prime}
I_2' &=& -{Q_2 \over 2
(-Q)^{1/2}} +{Q'' \over 8 (-Q)^{3/2}} +{5 \over 32} {Q'^{2} \over
(-Q)^{5/2}} \continue &=& -{Q_2 \over 2 (-Q)^{1/2}} +{Q'' \over 48
(-Q)^{3/2}} +{5 \over 48} \frac{d}{dx} \left({Q' \over (-Q)^{3/2}}
\right) \eea 
as follows from \refeq{eq:2ndTransp} with
$I_0'=+(-Q)^{1/2}$ and $I_1'= i  (\log( -Q))'/4$ after some
calculation. But, when this is integrated from $x$ up to $a-\mu$ (with $\mu \rightarrow 0+$) the last
two terms diverge algebraically with respect to $\mu$.

We turn to  the region near the turning point. The oscillatory solution is written in complex form as 
\beq
w_{osc}(x) \equiv  C \exp \left(i \int_{x}^{a-\mu} ds \, I'(s) \right)  
\eeq
We shall expand in the distance from the turning point
\beq
\xx = x-a 
\eeq
with $\xx <0$ corresponding to the oscillatory region.
Thus from \refeq{eq:PotExp}
\beq
-Q  = \axx  \, \alpha \left(1-\frac{\beta}{\alpha} \axx \right) \,.
\eeq
The leading action $I_0$ is found from
\beq
(-Q)^{1/2} \sim \axx^{1/2} \alpha^{1/2} \left(1-\frac{\beta}{2 \alpha} \axx \right) 
\eeq
so
\beq
\int_\xx^0  ds \, (-Q)^{1/2} \sim \alpha^{1/2} \left({2 \over 3} \axx^{3/2} - {\beta \over 5 \alpha^{1/2}} \axx^{5/2} \right) 
\eeq
The next action $I_1$ is imaginary and contributes to the amplitude of the JWKB-function with the factor 
\beq
(-Q)^{-1/4} \sim \axx^{-1/4} \, \alpha^{-1/4} \left(  1+ {1 \over 4}{\beta \over \alpha}  \axx \right) \,.
\eeq
The second order action $I_2$ follows from integration of \refeq{eq:I2Prime}. 
First,
\beq
{Q'' \over 48 (-Q)^{3/2}} \sim \frac{\beta}{24 \alpha^{3/2}} \axx^{-3/2}
\eeq
so
\beq
\int_x^{a-\mu} ds \,  \left( {Q'' \over 48 (-Q)^{3/2}} \right) \sim \frac{\beta}{12 \alpha^{3/2}} (\mu^{-1/2} -\axx^{-1/2}) \,. 
\eeq
Second,  the exact term given by a derivative becomes
\beq
{5 \over 48} \int_x^{a-\mu} ds \,  \frac{d}{ds} \left({Q' \over (-Q)^{3/2}} \right) \sim \frac{5}{48} \alpha^{-1/2}  \left(\mu^{-3/2} -\axx^{-3/2} \right) \,.
\eeq
Finally the subdominant potential term with $Q_2 = \gamma + O(\xx) $ (not having a zero at the turning point) becomes
\beq
-{Q_2 \over 2 (-Q)^{1/2}}  \sim -{\gamma \over 2 \alpha^{1/2}} \axx^{-1/2} 
\eeq
and leads to the regular
\beq
  \int_x^{a-\mu} ds \left( -{Q_2 \over 2 (-Q)^{1/2}} \right) \sim {\gamma \over \alpha} (\axx^{1/2} -\mu^{1/2}) \,. 
\eeq

On the other
hand, in the transition region $x\approx a$ with $x<a$ there is also a solution $w_{trans}$ in
the form of the Airy function (\cite{BenderOrzag}: formula (10.7.7)).
The Airy form is recognized in \refeq{eq:ODEbasic} by the transformation 
\beq \label{eq:Substitution}
x-a = \epsilon^{2/3} \alpha^{-1/3} t \,.
\eeq
Then a perturbed Airy equation arises
\beq
\frac{d^2 w_{trans}}{dt^2} = (t + \epsilon^{2/3} \alpha^{-4/3} \beta t^2) w_{trans} \,.
\eeq
The leading perturbation due to $Q$ is included whereas that from  $Q_2$ does not play a role at this order.
Consequently
\beq
w_{trans} \sim D \left(1-\frac{\beta x}{5 \alpha}\right) \mathrm{Ai} \left(\alpha^{1/3} \epsilon^{-2/3}  \left(x + \frac{\beta x^2}{5 \alpha} \right) \right) \,.
\eeq
We now expand this Airy function in the region $x<a$ to second order in an asymptotic series of oscillatory form:
\beq
\fl \mathrm{Ai}(-z) \sim \sin\left(\zeta + {\pi \over 4}\right) -\frac{5}{48} z^{-3/2} \cos\left(\zeta + {\pi \over 4}\right) \approx \sin\left(\zeta + {\pi \over 4} -\frac{5}{48} z^{-3/2}\right)
\eeq
for large $z$  with 
\beq
z = \alpha^{1/3} \epsilon^{-2/3}  \axx \left(1 - \frac{\beta \axx}{5 \alpha} \right) 
\eeq
and
\beq
\zeta = \frac{2}{3} \, z^{3/2} \sim {2 \over 3} {\alpha^{1/2}   \epsilon^{-1}} \axx^{3/2}  \left(1-{3 \beta \axx \over 10 \alpha} \right) \,.
\eeq
Calculations similar  as those for $w_{osc}$ give
\bea \fl
w_{trans}(x) \sim & & \continue 
& &\fl  \Im \left[ {D}{\sqrt{\pi}} \left( 1+ {\beta \axx \over 4 \alpha}\right)  \epsilon^{1/6} \alpha^{-1/12} \axx^{-1/4}   \exp\left( i \left(  {2 \over 3} \alpha^{1/2} \epsilon^{-1} \left( \axx^{3/2}-{3 \beta \axx^{5/2} \over 10 \alpha }\right)   \right. \right. \right.   \continue 
& &  \fl \left. \left. \left. +{\pi \over 4}  - {5 \over 48} \axx^{-3/2} \alpha^{-1/2}  \epsilon \left( 1 + {3 \beta \axx \over 10 \alpha} \right)  \right) \right)    \right] \,.
\eea
This result for $w_{trans}$ is  matched with the previous oscillatory form $w_{osc}$. First, we note that provided we neglect terms of order $\epsilon \axx^{-1/2}$ there is a complete match in functional form with respect to $\axx$ in the oscillatory region close to the turning point. In particular, $Q_2$ does not play a role in the matching at this order. Second, as $w_{trans}$ is independent of $\mu$  all dependence on $\mu$ must be removed in $w_{osc}$. This is done 
by adjusting the pre-factor
$C$ of the standard JWKB-solution $w_{osc}={\mathrm{Im}}(C \exp(i I))$.   In this way  singular terms in $\mu$ are absorbed. Thus for
example also the last term's singularity in \refeq{eq:I2Prime} proportional to $\mu^{-3/2}$ is removed. 

Still in our case, the contribution from the subdominant
$Q_2$ to the second correction to the phase is regular and eventually gives a
correction to the result stated in \cite{BenderOrzag} for the full phase (i.e. all the way to the boundary $x_0$): 
\beq
\label{eq:DeltaI2}
\Delta I_2 = -{1 \over 2} \int^a_{x_0}
{Q_2(s)\over I_0'(s)}ds=-{1 \over 2} \int^a_{x_0} {Q_2(s)\over
  (-Q(s))^{1/2}}ds \,.  \eeq
Likewise the integration of the remaining terms in \refeq{eq:I2Prime} are seen to correspond to those given in \refeq{eq:JWKBCorr}.

By construction the phase is now correct to first order in $\epsilon$.
This can be verified at the point $x=0$ as follows:
There the phase $I_2$ given by \refeq{eq:I2Short} evaluates  to 
\beq
I_2 = {\pi \over 16} \,.
\eeq
However, using 
\beq
\epsilon = l^{-1}-\frac{1}{2} \, l^{-2} + O(l^{-3}) \quad \mbox{and} \quad 1/\epsilon = l+\frac{1}{2}-\frac{1}{8} \, l^{-1} + \frac{1}{16} \, l^{-2}+ O(l^{-3})
\eeq
and \refeq{eq:I0at0} as well as \refeq{eq:AorPsi} on the oscillatory part
\beq \eqalign{
\psi &\propto \cos\left({I_0 \over \epsilon} + I_2 \epsilon-{\pi \over 4} \right) =\cos\left(\frac{\pi}{2}\, (l-m) +O(l^{-3})\right) \\
&\propto Y_{l m}(\theta=\pi/2,0)+O(l^{-3})}
\eeq
shows the agreement in the phase even to second order in $l^{-1}$. For further numerical checks see \ref{app:ScatPhNumeric}.

\subsubsection{Amplitude}
 
Even though  our main interest is a JWKB-resonance condition which only depends on the phase we  note that similar calculations are possible for the amplitude function. These show agreement between the JWKB result and the
exact to first order $l^{-1}$.

\subsubsection{JWKB-phase at eigenmodes} 
\label{app:ScatPhNumeric}
The  \tref{tab:wkbPh} (opening angle $\theta_0=\pi/3$) shows the deviation  of the  JWKB-scattering phase $\Theta$ measured in units of $2 \pi$ from the nearest integer (denoted $\left[\Theta/2\pi \right]$) normalized with that integer, i.e. 
\beq \label{eq:ScatPhDevMeas}
\frac{\Theta/2 \pi- \left[\Theta/2 \pi \right]}{\left[\Theta/2 \pi\right]} \,.
\eeq
Apart from the  value at $(m,l)=(13,31.74)$ an overall improvement is observed.

\begin{flushleft}
\begin{table}[ht]
\caption{\label{tab:wkbPh}  Scattering phase in units of $2 \pi$: the relative deviation from integer \eref{eq:ScatPhDevMeas} at $\theta_0 = \pi/3$.}
\item[]\begin{tabular}{llrl|llrl} \mr
$m$ &$ l $&$ \mathrm{Leading} $&$ \mathrm{Corrected} $&$ m $&$ l $&$ \mathrm{Leading} $&$ \mathrm{Corrected}$ \\  \br
$ 0 $&$ 1.777 $&$ -9.43\times 10^{\mathrm{-3}} $&$ -1.01\times 10^{\mathrm{-3}} $&$ 8 $&$ 21.94 $&$ -5.77\times 10^{\mathrm{-5}} $&$
   -1.81\times 10^{\mathrm{-6}} $\\
 $0 $&$ 19.75 $&$ -1.32\times 10^{\mathrm{-4}} $&$ -2.93\times 10^{\mathrm{-7}} $&$ 8 $&$ 40.31 $&$ -3.73\times 10^{\mathrm{-5}} $&$
   -4.37\times 10^{\mathrm{-8}} $\\
 $0 $&$ 37.75 $&$ -3.76\times 10^{\mathrm{-5}} $&$ -2.36\times 10^{\mathrm{-8}} $&$ 9 $&$ 26.40 $&$ -5.28\times 10^{\mathrm{-5}} $&$
   -7.20\times 10^{\mathrm{-7}} $\\
 $1 $&$ 3.196 $&$ -3.79\times 10^{\mathrm{-3}} $&$ -6.68\times 10^{\mathrm{-4}} $&$ 10 $&$ 18.14 $&$ 6.75\times 10^{\mathrm{-4}} $&$
   -3.30\times 10^{\mathrm{-5}} $\\
 $1 $&$ 21.24 $&$ -1.22\times 10^{\mathrm{-4}} $&$ -2.45\times 10^{\mathrm{-7}} $&$ 10 $&$ 37.00 $&$ -3.95\times 10^{\mathrm{-5}} $&$
   -1.05\times 10^{\mathrm{-7}} $\\
 $1 $&$ 39.24 $&$ -3.61\times 10^{\mathrm{-5}} $&$ -2.12\times 10^{\mathrm{-8}} $&$ 11 $&$ 29.08 $&$ -2.42\times 10^{\mathrm{-5}} $&$
   -7.19\times 10^{\mathrm{-7}} $\\
 $2 $&$ 7.622 $&$ -8.31\times 10^{\mathrm{-4}} $&$ -3.39\times 10^{\mathrm{-5}} $&$ 12 $&$ 20.68 $&$ 8.30\times 10^{\mathrm{-4}} $&$
   -3.29\times 10^{\mathrm{-5}} $\\
 $2 $&$ 25.71 $&$ -8.69\times 10^{\mathrm{-5}} $&$ -1.30\times 10^{\mathrm{-7}} $&$ 12 $&$ 39.74 $&$ -3.06\times 10^{\mathrm{-5}} $&$
   -1.05\times 10^{\mathrm{-7}} $\\
$ 2 $&$ 43.73 $&$ -2.99\times 10^{\mathrm{-5}} $&$ -1.46\times 10^{\mathrm{-8}} $&$ 13 $&$ 31.74 $&$ 4.69\times 10^{\mathrm{-8}} $&$
   -7.18\times 10^{\mathrm{-7}} $\\
 $3 $&$ 8.991 $&$ -4.80\times 10^{\mathrm{-4}} $&$ -3.35\times 10^{\mathrm{-5}} $&$ 13 $&$ 34.88 $&$ -1.89\times 10^{\mathrm{-5}} $&$
   -3.39\times 10^{\mathrm{-7}} $\\
 $3 $&$ 27.16 $&$ -8.00\times 10^{\mathrm{-5}} $&$ -1.21\times 10^{\mathrm{-7}} $&$ 15 $&$ 24.46 $&$ 1.01\times 10^{\mathrm{-3}} $&$
   -3.27\times 10^{\mathrm{-5}} $\\
 $3 $&$ 45.20 $&$ -2.87\times 10^{\mathrm{-5}} $&$ -1.37\times 10^{\mathrm{-8}} $&$ 16 $&$ 21.99 $&$ 6.61\times 10^{\mathrm{-3}} $&$
   -6.23\times 10^{\mathrm{-4}} $\\
 $4 $&$ 19.53 $&$ -1.51\times 10^{\mathrm{-4}} $&$ -7.47\times 10^{\mathrm{-7}} $&$ 17 $&$ 23.21 $&$ 6.72\times 10^{\mathrm{-3}} $&$
   -6.22\times 10^{\mathrm{-4}} $\\
 $4 $&$ 37.64 $&$ -4.23\times 10^{\mathrm{-5}} $&$ -3.33\times 10^{\mathrm{-8}} $&$ 18 $&$ 24.42 $&$ 6.82\times 10^{\mathrm{-3}} $&$
   -6.21\times 10^{\mathrm{-4}} $\\
 $5 $&$ 17.87 $&$ -1.58\times 10^{\mathrm{-4}} $&$ -1.83\times 10^{\mathrm{-6}} $&$ 19 $&$ 29.45 $&$ 1.17\times 10^{\mathrm{-3}} $&$
   -3.25\times 10^{\mathrm{-5}} $\\
$ 5 $&$ 36.06 $&$ -4.67\times 10^{\mathrm{-5}} $&$ -4.65\times 10^{\mathrm{-8}} $&$ 20 $&$ 34.22 $&$ 3.81\times 10^{\mathrm{-4}} $&$
   -5.93\times 10^{\mathrm{-6}} $\\
$ 6 $&$ 16.14 $&$ -9.84\times 10^{\mathrm{-5}} $&$ -6.02\times 10^{\mathrm{-6}} $&$ 21 $&$ 38.87 $&$ 1.59\times 10^{\mathrm{-4}} $&$
   -1.80\times 10^{\mathrm{-6}} $\\
 $6 $&$ 34.47 $&$ -5.11\times 10^{\mathrm{-5}} $&$ -6.86\times 10^{\mathrm{-8}} $&$ 23 $&$ 34.40 $&$ 1.29\times 10^{\mathrm{-3}} $&$
   -3.23\times 10^{\mathrm{-5}} $\\
 $7 $&$ 20.59 $&$ -8.76\times 10^{\mathrm{-5}} $&$ -1.81\times 10^{\mathrm{-6}} $&$ 25 $&$ 36.86 $&$ 1.34\times 10^{\mathrm{-3}} $&$
   -3.22\times 10^{\mathrm{-5}} $\\
 $7 $&$ 38.91 $&$ -4.03\times 10^{\mathrm{-5}} $&$ -4.44\times 10^{\mathrm{-8}} $&$ 28 $&$ 36.46 $&$ 7.48\times 10^{\mathrm{-3}} $& 
$   -6.14\times 10^{\mathrm{-4}}$ \\ \mr
\end{tabular}
\end{table}
\end{flushleft}

\section{Scattering states}
\label{app:ScatState}
\subsection{Exact scattering states} 
One possible set of exact scattering states  for the sphere $\mathcal{S}^2$ are those used
extensively in geophysics under the name {\it traveling waves} 
in the discussion of the Green's function
and associated surface waves \cite{DT98}. We define them normalized:
\beq \label{defScatCurv} \psi^{(\pm) m}_l =
\sqrt{\frac{\Gamma(l-m+1)}{\Gamma(l+m+1)}} \, \left(P_l^m(x) \mp i
\frac{2}{\pi} Q_l^m(x) \right) \, \exp(i m \phi) \,.  \eeq 
The opposite sign in \refeq{defScatCurv} will be explained in the
 following.

With this choice each state
has constant total flux $F$ through circles of constant $\theta$: 
\beq F = \int_{\theta = cst} \bm{J \cdot \hat{\theta}}\,
ds = \int J_{\hat{\theta}} \, ds \eeq with $ds = R \sin \theta \,
d\phi$. The corresponding probability current is proportional to
(ignoring factors of $\hbar$ and mass): \beq J_{\hat{\theta}} =
\frac{1}{2 i}\,\left(\psi^{(+) m *}_l \, \partial_{\hat{\theta}}
\psi^{(+) m}_l - c.c. \right) \,.  \eeq The directional derivative is
rewritten as \beq \partial_{\hat{\theta}} = \frac{1}{R} \,
\frac{\partial}{\partial \theta} = -\frac{\sin \theta}{R} \,
\frac{\partial}{\partial x} \, , \eeq where $x=\cos \theta$.  The
current in terms of the Wronskian becomes \beq J_{\hat{\theta}} =
\frac{2}{\pi \, R}\, \frac{\Gamma(l-m+1)}{\Gamma(l+m+1)}
\,Wr(P_l^m,Q_l^m) \, \sin \theta \eeq and hence the flux \beq F= 4 \,
\sin^2 \theta \, Wr(P_l^m,Q_l^m)\,
\frac{\Gamma(l-m+1)}{\Gamma(l+m+1)}\,.  \eeq Using the Wronskian
\cite{MO53} (p.145: formula (25) in chapter 3.4 for Legendre functions
on the {\it cut}: $-1 \leq x \leq 1$ ): \beq
\label{eq:Wr}
Wr(P_l^m(x),Q_l^m(x)) =\frac{2^{2 m}
\Gamma(\frac{l+m+2}{2})\Gamma(\frac{l+m+1}{2})}{(1-x^2)
\Gamma(\frac{l-m+2}{2}) \Gamma(\frac{l-m+1}{2})} \eeq with the
duplication formula for the Gamma function \refeq{eq:GammaDupl} for
the numerator and denominator in \eref{eq:Wr} then gives \beq F =
4>0 \, \eeq corresponding to an {\it outgoing} state. The constancy of
the flux reflects probability conservation.

\subsection{Semiclassical scattering states}
As in \ref{app:JWKB} we  match the JWKB traveling states to the exact scattering states with the result
\beq
(a^2-x^2)^{-1/4} \exp\left(\pm i \left({I_0 \over \epsilon}-{\pi \over 4}\right)\right) \approx (-1)^m \sqrt{2 l+1 \over 4 \pi} \, \pi \,  \psi^{(\pm) m}_l
\eeq
using \eref{eq:Leg1at0} and 
\beq
Q_l^m(0)=-{2^{m-1}} \,{\pi^{1/2}} \,
\frac{\Gamma{(\frac{l+m+1}{2}})}{\Gamma{(\frac{l-m+2}{2}})} \,
\sin\left(\frac{\pi}{2} (l+m)\right) \,.
\eeq
Hence, the scattering states lead to  Dirichlet scattering phases given
asymptotically by 
\beq
-\frac{ \psi^{(-) m}_l}{ \psi^{(+) m}_l} \approx \exp\left(-i\left({2 I_0 \over \epsilon} +{\pi \over 2} \right)\right) = \exp(-i \Theta_m) \equiv S_m
\eeq
with $\Theta_m$ given by \eref{eq:SingleScatPhase} to leading order.

\end{document}